\documentclass[aps,twocolumn,showpacs]{revtex4}

\usepackage{mathptm}    %clears up some problems for generating PDF file
\usepackage{dcolumn}                    % Align table columns on decimal point
\usepackage{bm}                         % bold math
\usepackage{bbm}
\usepackage{dsfont}
\usepackage{graphicx}
\usepackage{amsmath}
\usepackage{feynmp}

\usepackage{appendix}
\usepackage{subfigure}

\newcommand {\ba} {\begin{eqnarray}}
\newcommand {\ea} {\end{eqnarray}}

\usepackage{times}

\begin{document}

\title{Pairing Mechanism of the Heavily Electron Doped FeSe Systems:\\ Dynamical Tuning of the Pairing Cutoff Energy}

\author{Yunkyu Bang}
\email[]{ykbang@chonnam.ac.kr}
\affiliation{Department of Physics,
Chonnam National University, Kwangju 500-757, Republic of Korea}

\begin{abstract}
We studied pairing mechanism of the heavily electron doped FeSe (HEDIS) systems, which commonly have one incipient hole band -- a band top below the Fermi level by a finite energy distance $\epsilon_b$ -- at $\Gamma$ point and
ordinary electron bands at $M$ points in Brillouin zone (BZ).
We found that the system allows two degenerate superconducting solutions with the exactly same $T_c$ in clean limit: the incipient $s^{\pm}_{he}$-gap ($\Delta_h^{-} \neq 0$, $\Delta_e^{+} \neq 0$) and $s_{ee}^{++}$-gap ($\Delta_h =0$, $\Delta_e^{+} \neq 0$) solutions with different pairing cutoffs, $\Lambda_{sf}$ (spin fluctuation energy) and $\epsilon_b$, respectively.
The $s_{ee}^{++}$-gap solution, in which the system dynamically renormalizes the original pairing cutoff  $\Lambda_{sf}$ to $\Lambda_{phys}=\epsilon_b$  ($< \Lambda_{sf}$), therefore actively eliminates the
incipient hole band from forming Cooper pairs, but without loss of $T_c$, becomes immune to the impurity pair-breaking.
As a result, the HEDIS systems, by dynamically tuning the pairing cutoff and therefore selecting the $s_{ee}^{++}$-pairing state, can always achieve the maximum $T_c$ -- the $T_c$ of the degenerate $s^{\pm}_{he}$ solution in the ideal clean limit -- latent in the original pairing interactions, even in dirty limit.
\end{abstract}


\date{\today}
\maketitle

\section{Introduction.}
The discovery of the FeSe/SrTiO$_3$ monolayer system ($T_c \approx 60-100K$) \cite{FeSe1,FeSe2,FeSe3} and other heavily
electron-doped iron selenide (HEDIS) compounds such as
A$_x$Fe$_{2-y}$Se$_2$ (A=K, Rb, Cs, Tl, etc.) ($T_c \approx
30-40K$)\cite{HEDIS1,HEDIS2,HEDIS3}, (Li$_{1-x}$Fe$_x$OH)FeSe ($T_c \approx
40K$) \cite{OHFeSe}, and pressurized bulk FeSe ($T_c \approx
37K$) \cite{pressure_FeSe} are posing a serious challenge to our
understanding of the Iron-based superconductors (IBS). The main
puzzles are two:\\
(1) why is $T_c$ so high, up $100K$ ? \\
(2) what is the pairing mechanism and pairing solution with
only electron pockets at $M$ point?

Among the HEDIS systems, we think, FeSe/SrTiO$_3$
monolayer system\cite{FeSe1,FeSe2,FeSe3} has one extra mechanism, which is the long-sought {\it small-angle
scattering phonon boost effect}\cite{Bang_phonon1,Bang_phonon2}.
Lee {\it et al.}\cite{FeSe_phonon} have measured the presence of
the ferroelectric polar phonon in the SrTiO$_3$(STO) substrate and
its strong coupling with the conduction band of the FeSe
monolayer. Subsequently, theoretical works\cite{DHLee_phonon,DHLee_phonon2,Johnston,Johnston2}
have elaborated this phonon boost effect specifically to the
FeSe-monolayer system.
This phonon boost effect is theoretically trivial to understand. When
there exist a large momentum exchange repulsive interaction $V_{Q}$
-- provided by antiferromagnetic(AFM) spin fluctuations -- and a
small momentum exchange attractive phonon interaction $V_{ph}$, two pairing
potentials do not interfere each other living in the different
sectors of momentum space but work together to boost $T_c$ of the
$s_{\pm}$- or $d$-wave gap solutions as following
way\cite{Bang_phonon1,Bang_phonon2}:
\begin{equation}
T_c \simeq 1.14~ \Lambda_{sf} ^{\tilde{\lambda}_{sf}} \cdot
\Lambda_{ph} ^{\tilde{\lambda}_{ph}} ~~e^{-1/\lambda_{tot}}
\end{equation}
where $\tilde{\lambda}_{sf} =\lambda_{sf} /\lambda_{tot}$,
$\tilde{\lambda}_{ph} =\lambda_{ph} /\lambda_{tot}$, and
$\lambda_{tot} =(\lambda_{sf} +\lambda_{ph})$. $\lambda_{sf,ph}$
and $\Lambda_{sf,ph}$ are the dimensionless coupling constants and
the characteristic energy scales of the spin fluctuations and
phonon, respectively. The Eq.(1) shows that the phonon coupling
$\lambda_{ph}$ -- even if weak strength -- entering into the exponent
of the exponential as $\lambda_{tot}=(\lambda_{sf} +\lambda_{ph})$, its
boosting effect of $T_c$ can be far more efficient than a simple
algebraic addition.

\begin{figure}
\noindent
\includegraphics[width=80mm]{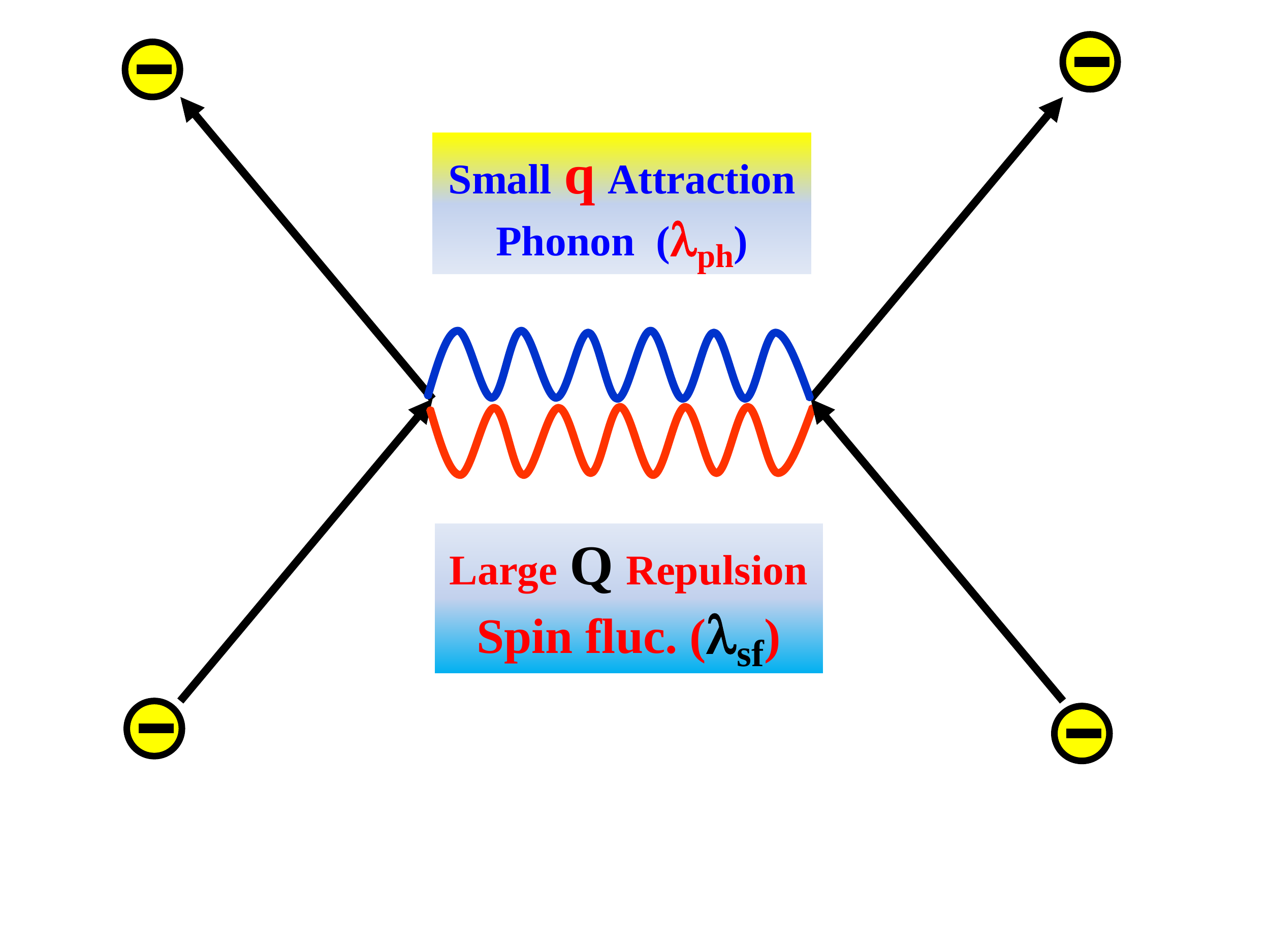}
\vspace{-1.0cm}
\caption{(Color online) Schematic Feynman diagram of the phonon boost effect. A small momenta $\vec{q}$ exchanging phonon attractive interaction $\lambda_{ph}$ and the large momenta $\vec{Q}$ exchanging AFM repulsive interaction $\lambda_{AFM}$, living in different momentum space, do not interfere but cooperate to enhance the total pairing interaction $\lambda_{tot}=\lambda_{AFM}+\lambda_{ph}$ for the $s^{\pm}$- and $d$-wave pairing channels.
\label{fig1}}
\end{figure}

Apart from this phonon boost effect, which exists -- or particularly strong -- only in the FeSe/STO monolayer system, theoretically more challenging question of the HEDIS superconductors is:
{\it even without the phonon boosting effect, how these systems can achieve such high $T_c$ of 30-40K only with the electron pockets}. To clarify this question, we would like to understand the following more specific questions:\\
(1)  without the hole pocket at $\Gamma$ point, what is the
pairing mechanism and the pairing state ? \\
(2) does the incipient (sunken) hole band have
any specific role or mechanism for such high $T_c$ ? \\
(3) why do all HEDIS compounds seem to have the optimal
incipiency distance $\epsilon_b^{optimal}$ of 60-80 $meV$ for the maximum
$T_c$ ?

In this paper, we provide a noble mechanism as an unified answer to all these questions, that is
{\it dynamical tuning of the pairing cutoff energy.}
We studied a simple model for the HEDIS systems which consists of one
incipient hole band and ordinary electron band(s) with a dominant
repulsive interband interaction $V_{inter}$ between them. The original pairing
interaction cutoff energy $\Lambda_{sf}$ is assumed to be larger than the
incipient energy $\epsilon_b$ ($< \Lambda_{sf}$). This model system has been recently
studied, and was shown to have the $s^{\pm}_{he}$-gap solution\cite{Bang_shadow, Hirschfeld_incipient}.
This gap solution forms non-zero gaps both on the incipient hole band and electron band ($\Delta_h \neq 0$, $\Delta_e \neq 0$), hence it is specifically named as "incipient" $s^{\pm}_{he}$-gap solution.
In this paper, we showed that this incipient band model allows another degenerate pairing solution with the same $T_c$ as the incipient $s^{\pm}_{he}$-gap solution, but only with the electron
bands ($\Delta_h =0$, $\Delta_e \neq 0$), hence named as $s_{ee}^{++}$-gap solution. For this pairing solution, the incipient hole band is dynamically eliminated by the renormalization
process, and the bare pairing interactions $V^{0}_{ab}(\Lambda_{sf})
>0 (a,b=h,e)$ get renormalized and in particular the bare repulsive intraband interaction ($V^{0}_{ee} (\Lambda_{sf}) >0$) for the electron band evolves into an attractive
interaction ($V^{ren}_{ee}(\Lambda=\epsilon_b) <0$) with a reduced pairing cutoff energy $\Lambda=\epsilon_b$,
with which the system can form a s-wave pairing with the electron bands only, namely, the $s_{ee}^{++}$-gap.

The idea of the s-wave pairing state ($s_{ee}^{++}$) in the electron bands only
was also proposed in the previous works\cite{Wang2013, DHLee2011} using the functional renormalization group (FRG) technique. Although the FRG approach correctly captured the leading instability of the $s_{ee}^{++}$-pairing state, this method could not distinguish the $s_{ee}^{++}$-state and the incipient-$s^{\pm}_{he}$-state, nor did it clarify the subtle relation between these two pairing states. The FRG techniques traces the RG flow of the leading pairing form factor $\phi_{\alpha}(k)$ by reducing the scaling energy $\Lambda$ until its eigenvalue diverges. However, the form factor $\phi_{\alpha}(k)$, being a lattice harmonics in the $C_4$ symmetric lattice, is the same both for the incipient $s^{\pm}_{he}$-state and the $s_{ee}^{++}$-state. Therefore when the eigenvalue of this form factor diverges at $\Lambda^{*}$, the FRG result just indicates $T_c \sim \Lambda^{*}$ in the pairing channel $\phi_{\alpha}$, so that it cannot determine which of two pairing states --the incipient $s^{\pm}_{he}$-state and the $s_{ee}^{++}$-state -- is the real ground state because the FRG technique itself does not determine the physical pairing cutoff $\Lambda_{phys}$.

Our important finding is that in the clean limit of the genuine incipient band model these two pairing states -- the incipient $s^{\pm}_{he}$-state  ($\Delta_h \neq 0$, $\Delta_e \neq 0$) and
$s_{ee}^{++}$-state  ($\Delta_h =0$, $\Delta_e \neq 0$) -- are degenerate two physical solutions with the exact same $T_c$ but with different physical pairing cutoff energies, $\Lambda_{phys}=\Lambda_{sf}$ and $\Lambda_{phys}=\epsilon_b$, respectively. These two states are both physical and distinct states.
On the one hand, our finding is in accord with the important principle of {\it the physical invariance of the renormalization group (RG) transformation}\cite{note1}. But on the other hand, our finding reveals a new potential of the RG transformation: the RG transformation can go beyond a mathematical technique to conveniently study the low energy physics\cite{note2} and the system can actively utilize its own RG flow to determine the physical cutoff energy scale $\Lambda_{phys}$ and optimize its best ground state. This active mechanism of RG is not new but already known with the renormalization of the Coulomb pseudopotential\cite{Schrieffer}. The degeneracy of these two pairing solutions of the HEDIS system becomes broken by impurities and the system choose the $s_{ee}^{++}$-gap solution and extracts the maximum $T_c$, potentially stored in the system by avoiding the impurity pair-breaking.

In section II, we illustrated the concept of this dynamical tuning of the cutoff energy scale with the well known example of the Coulomb pseudopotential and the phonon-mediated BCS superconductor\cite{Schrieffer}. We then showed that even the phonon-mediated BCS superconductor with a phonon energy $\omega_D$ needs not to have the physical cutoff $\Lambda_{phys}$ to be $\omega_D$, as commonly believed, but can take arbitrary scale $\Lambda$ without affecting $T_c$, $\Delta_{sc}$, and the condensation energy, within the RG scheme. And higher order correction is necessary to determine the physical cutoff as $\Lambda_{phys}=\omega_D$.

In section III, equipped with this new concept of dynamical tuning of cutoff energy by RG,  we studied the incipient band model with one incipient hole band with $\epsilon_b$ and electron band(s) mediated by dominant interband repulsive potential $V_{inter-band} > V_{intra-band}$ with original pairing cutoff $\Lambda_{sf} > \epsilon_b$. In subsection III.A, first we studied the minimal two band model and demonstrated that the $T_c$ is invariant with scaling $\Lambda < \Lambda_{sf}$. Therefore when the scaling energy $\Lambda$ crossovers from above to below $\epsilon_b$, the pairing solution continuously changes from the incipient $s^{\pm}_{he}$-state to the $s_{ee}^{++}$-state keeping the same symmetry and same $T_c$. In section III.B, we showed that the non-magnetic impurity scattering severely weakens the incipient $s^{\pm}_{he}$-state, but would not affect the $T_c$ of the $s_{ee}^{++}$-state with the physical cutoff $\Lambda_{phys}=\epsilon_b$.

We propose that this is the key mechanism why the HEDIS systems can achieve reasonably high $T_c$ of 30-40K with the sunken (incipient) hole band. They can avoid the impurity pair-breaking by dynamical tuning the pairing cutoff energy to $\epsilon_b$.
The standard IBS systems with both hole and electron bands crossing the Fermi level cannot have the choice of the $s_{ee}^{++}$-solution but only the $s^{\pm}_{he}$-solution, therefore the standard IBS systems would suffer severe reduction of $T_c$ from the inevitable impurities introduced by dopings. Otherwise all standard IBS systems could have achieved much higher $T_c$. On the other hand, the $s_{ee}^{++}$-solution with the sunken hole band has a different drawback; i.e. increasing the incipient distance $\epsilon_b$ weakens the pair susceptibility, hence reduces $T_c$. Therefore, we could expect that the maximum $T_c$ of the $s_{ee}^{++}$-solution should occur with $\epsilon_b \rightarrow 0$, but it is not the case in experiments.
We found that there exists a mechanism to determine an optimal value of $\epsilon_b^{optimal}$.
The above mentioned RG scaling breaks down when $\epsilon_b$ becomes too small because the pair susceptibility for the $s_{ee}^{++}$-state, $\chi(\epsilon_b)=-2 T_c\sum_n \int^{\epsilon_b}_{0} d\epsilon \frac{1}{\omega_n^2 + \epsilon^2(k)}$, ($\omega_n=\pi T_c(2n+1))$, becomes saturated as $\epsilon_b \rightarrow T_c$, which sets the stable minimum cutoff energy scale $\Lambda_{phys}= \epsilon_b \approx T_c$. In real system, the optimal $\Lambda_{phys}$ should increase further by other broadening processes as $\Lambda_{phys} = \epsilon_b^{optimal} \approx (\pi T_c + \Gamma_{imp} +\Gamma_{inela})$ (where $\Gamma_{imp}$ is the impurity scattering rate and $\Gamma_{inelas}$ is the inelastic scattering rate). This is the reason why optimal incipient energy $\epsilon_b^{optimal}$ is about 60-80 $meV$ in all HEDIS systems.

In section III.C and D, we studied a more realistic three band model with one incipient hole band and two electron bands $e1$ and $e2$. With this model, we could consider another possible pairing solution: the $s^{+-}_{e1e2}$-state, also called, "nodeless" $d$-wave state, in which $\Delta_{e1}=-\Delta_{e2}$\cite{d-wave1, d-wave2, d-wave3}.
We showed in general that the $s^{+-}_{e1e2}$-state is favored when the incipient hole band is deep (larger $\epsilon_b$) but the $s^{++}_{e1e2}$-state becomes winning when the incipient hole band becomes intermediate to shallow (optimal $\epsilon_b$). With non-magnetic impurity scattering, we showed that the $s^{+-}_{e1e2}$-state (nodeless $d$-wave) is most rapidly destroyed, but the $s^{++}_{e1e2}$-state is immune to it and can survive with high $T_c$ in the region of optimal values of $\epsilon_b^{optimal}$. On the other hand, the "incipient" $s^{-++}_{he1e2}$-state -- which had the same $T_c$ as the $s^{++}_{e1e2}$-state in clean limit -- can survive with much reduced $T_c$ in the region of small values of $\epsilon_b$ if the impurity scattering is not strong enough to completely kill this pairing state. This double-dome structure of the phase diagram (see Fig.10(B)) is quite similar with recent experiments of electron doped FeSe systems\cite{double_dome1,double_dome2,double_dome3,FeSe_dosing,FeSe_dosing2}.

\section{Renormalization of Pairing Interactions}

The bare Coulomb repulsion $\mu_0$, operating up to the plasma frequency $\omega_{pl}$, is renormalized to become pseudopotential $\mu^{*}$, operating up to the phonon frequency $\omega_{D} ( < \omega_{pl})$, as follows\cite{Schrieffer},
\begin{equation}
\mu^{*}(\omega_{D}) = \mu_0 + \mu_0 \cdot \chi(\omega_{pl};\omega_D) \cdot \mu^{*}.
\end{equation}
Using Cooperon propagator $\chi(\omega_{pl};\omega_{D})=-2 T \sum_n N_0 \int^{\omega_{pl}}_{\omega_{D}} d\epsilon \frac{1}{\omega_n^2 + \epsilon^2(k)} \approx - N_0\ln[\frac{\omega_{pl}}{\omega_{D}}]$, we obtain the well known result,
\begin{equation}
\mu^{*}(\omega_{D}) = \frac{\mu_0}{1+\mu_0 N_0 \ln[\frac{\omega_{pl}}{\omega_{D}}]},
\end{equation}
where $N_0$ is the density of states (DOS) at Fermi level. Since $\omega_{pl} \gg \omega_{D}$, the strong repulsive Coulomb potential $\mu$ becomes much weakened Coulomb pseudo-potential $\mu^{*} \ll \mu$. Therefore the total BCS pair potential $V_{pair}(\omega_D)=[V_{ph}+\mu^{*}] (<0)$ can now be attractive with the common pairing cutoff energy $\Lambda_{phys}=\omega_{D}$. This is the well known mechanism of how weak phonon attraction $V_{ph} (<0)$ can overcome the stronger Coulomb repulsion $\mu_0 (>0)$ by retardation ($\omega_{D} \ll \omega_{pl}$). But the important message for us is that the physical pairing cutoff is not necessarily fixed by the boson energy scale of the corresponding interaction (in this example, $\omega_{pl}$).

A standard theory of the renormalization of the pairing interactions stops here.
But now we would like to perform a thought-experiment, namely, we continue to scale down the effective cutoff energy $\Lambda$ below $\omega_{D}$ to see what happens. It is straightforward to
continue the RG scaling as
\begin{equation}
V_{pair}(\Lambda) = \frac{V_{pair}(\omega_D)}{1+V_{pair}(\omega_D)N_0\ln[\frac{\omega_{D}}{\Lambda}]}.
\end{equation}
For an attractive interaction ($V_{pair} <0$), Eq.(4) indicates that the strength of $|V_{pair}(\Lambda)|$
increases as the cutoff energy $\Lambda$ decreases $(\Lambda < \omega_{D})$.
But it is straightforward to show that the $T_c$ is invariant with RG flow as
$T_c=1.14 \omega_D e^{-1/N_0|V_{pair}(\omega_D)|}=1.14 \Lambda e^{-1/N_0|V_{pair}(\Lambda)|}$. Furthermore, as far as the BCS limit ($\Delta_{sc}/\Lambda \ll 1$) holds, the superconducting (SC) gap size $\Delta_{sc}$ is invariant as $\Delta_{sc}(\omega_{D})=\Delta_{sc}(\Lambda)$ and the total condensation energy $\Delta E =\frac{1}{2} N_0 \Delta_{sc}(\Lambda)$ is also invariant with respect to the RG scaling. Therefore all physical quantities are not affected by this level of RG scaling.

Therefore, at this level, the system has no reason to choose $\omega_{D}$, the Debye frequency, as a physical pairing cutoff energy and higher order corrections in $O(\Delta_{sc}/\Lambda)$ are necessary to determine the true physical cutoff energy $\Lambda_{phys}$. Exact calculations of all higher order corrections are difficult but a simple hint comes from the total condensation energy expression, more precise form of which is given as $\Delta E = -N_0 \Lambda \sqrt{\Lambda^2 +\Delta_{sc}^2} + N_0 \Lambda^2 \approx -\frac{1}{2}N_0 \Delta_{sc}^2 +\frac{1}{8}N_0 \Delta_{sc}^2 (\frac{\Delta_{sc}}{\Lambda})^2 + \ldots$\cite{Bang_CE}. The second term is the next order correction to the BCS approximation $\Delta E_{BCS}=-\frac{1}{2}N_0 \Delta_{sc}^2$ and tells us that more condensation energy is gained with larger cutoff energy $\Lambda$ when the gap size $\Delta_{sc}$ is the same.
Therefore, among the degenerate $T_c(\Lambda)$ solutions, the system would choose the largest cutoff solution that is $\Lambda_{phys}=\omega_{D}$ in this particular case.

With the above exercises we would like to propose the key concept of this paper, i.e., {\it the physical pairing cutoff energy $\Lambda_{phys}$ of the SC transition is not automatically determined by the physical boson energy scales of the pairing interactions such as $\omega_{pl}$, $\omega_{D}$, or $\omega_{sf}$. But it can be dynamically tuned by the system to maximize the $T_c$ and the condensation energy.}

\section{Incipient Band Model}

In this section, we would like to apply the concept of {\it "dynamical tuning of cutoff energy"} discussed in the previous section to the incipient band model for the HEDIS systems.
At the moment, the most accepted theory of the superconductivity for the Fe-pnictide superconductors is the sign-changing s-wave pairing state ($s^{\pm}_{he}$) between hole band(s) around $\Gamma$ and electron band(s) around $M$ points in BZ, mediated by the antiferromagnetic (AFM) fluctuations with the wave vector ${\bf Q}$  connecting $\Gamma$ and $M$ points (C-type)\cite{Hirschfeld, Hirschfeld2}. However, the HEDIS systems commonly have only electron pockets(s) at $M$ points and the hole pocket is missing at $\Gamma$ point, which exists only as an incipient hole-band (see Fig.2). However, even without the hole pocket, experimental evidences are that the AFM spin correlation is dominantly the C-type with the characteristic wave vector $\vec{Q}$ connecting the incipient hole band and the electron band\cite{C-type}. Therefore, it is a natural attempt to extend the standard paradigm of the $s^{\pm}_{he}$ pairing mechanism to the HEDIS systems: a dominant pairing interaction, $V_{he} >0$, between the incipient hole band and electron band(s). There is possibly some fraction of deviation from C-type AFM correlation toward G-type AFM correlation with the wave vector ${\bf Q^{'}}$ connecting two electron bands at ${\bf Q_x}=(\pi,0)$ and ${\bf Q_y}=(0,\pi)$\cite{G-type,G-type2,G-type3}. Phenomenologically, we will consider this deviation by introducing another weaker repulsive interaction between two electron bands $V_{e1e2} >0$.

\begin{figure}
\noindent
\includegraphics[width=90mm]{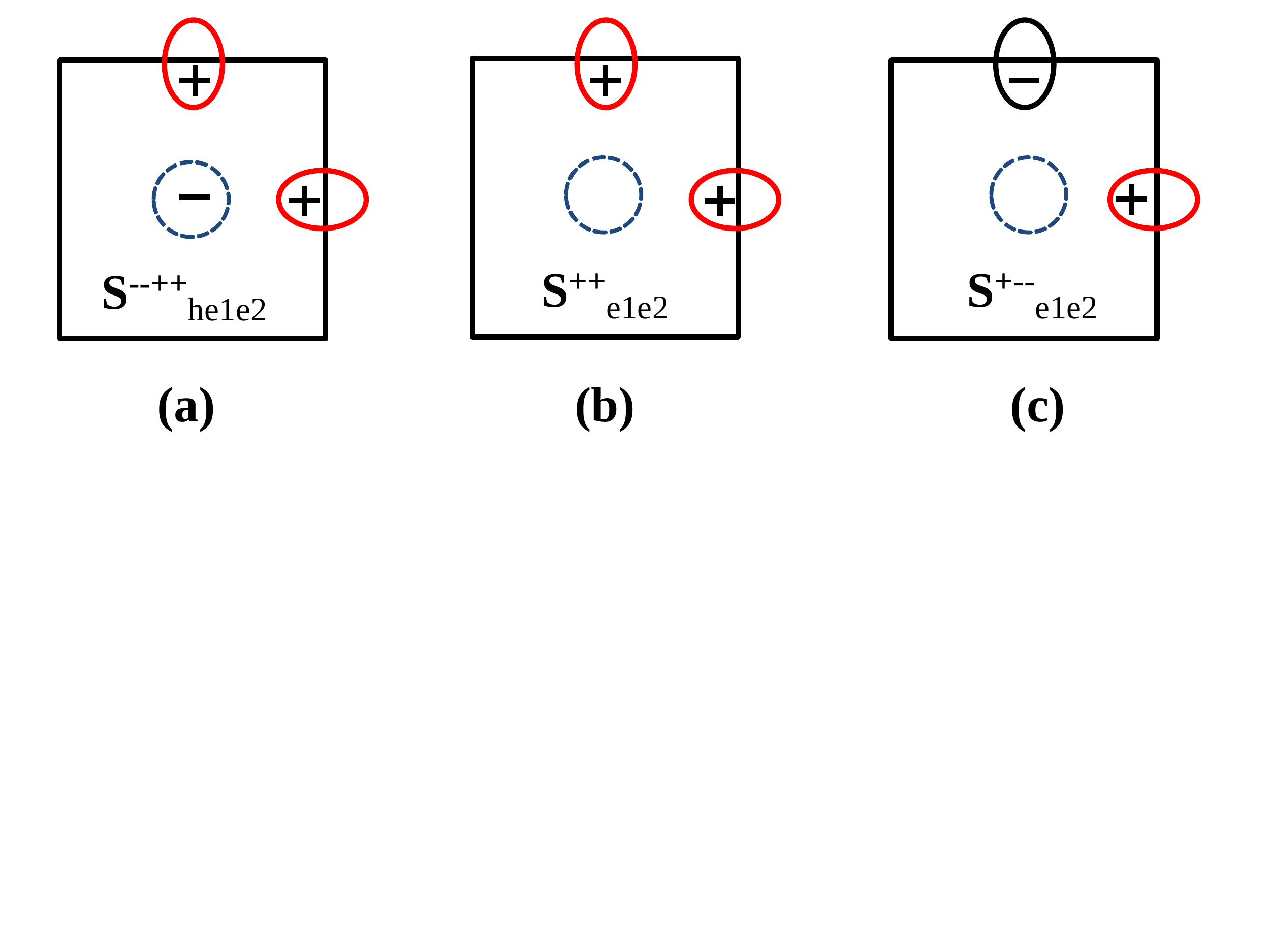}
\vspace{-4cm}
\caption{(Color online) Schematic pictures of three possible SC gap solutions of the incipient band model. In all three cases, the incipient (sunken) hole band is depicted as a dotted line circle around $\Gamma$ point. (a) incipient $s_{he}^{-+}$-gap solution. (b) $s^{++}_{e1e2}$-gap solution with no Cooper pairs formed on the incipient hole band. (c) $s^{+-}_{e1e2}$-gap solution with no Cooper pairs formed on the incipient hole band, also called as "nodeless $d$-wave".
\label{fig2}}
\end{figure}

\subsection{Incipient Two Band Model}

In order to illustrate the essence of the dynamical tuning mechanism and RG scaling of the cutoff energy, we first study a minimal two band model with one incipient hole band and one electron band as depicted in Fig.3(a). Here we ignore the interaction between electron bands $e1$ and $e2$, therefore two electron bands, $e1$ and $e2$, can be considered as identical one e-band as far as the SC pairing mechanism is concerned. As a result, among the three possible pairing states depicted in Fig.2, only $s_{he}^{-+}$-gap (Fig.2(a)) and $s^{++}_{ee}$-gap (Fig.2(b)) are possible with the two band model. This simplification is only for the proof of concept and we will consider the full three band model later.

\begin{figure}
\noindent
\includegraphics[width=95mm]{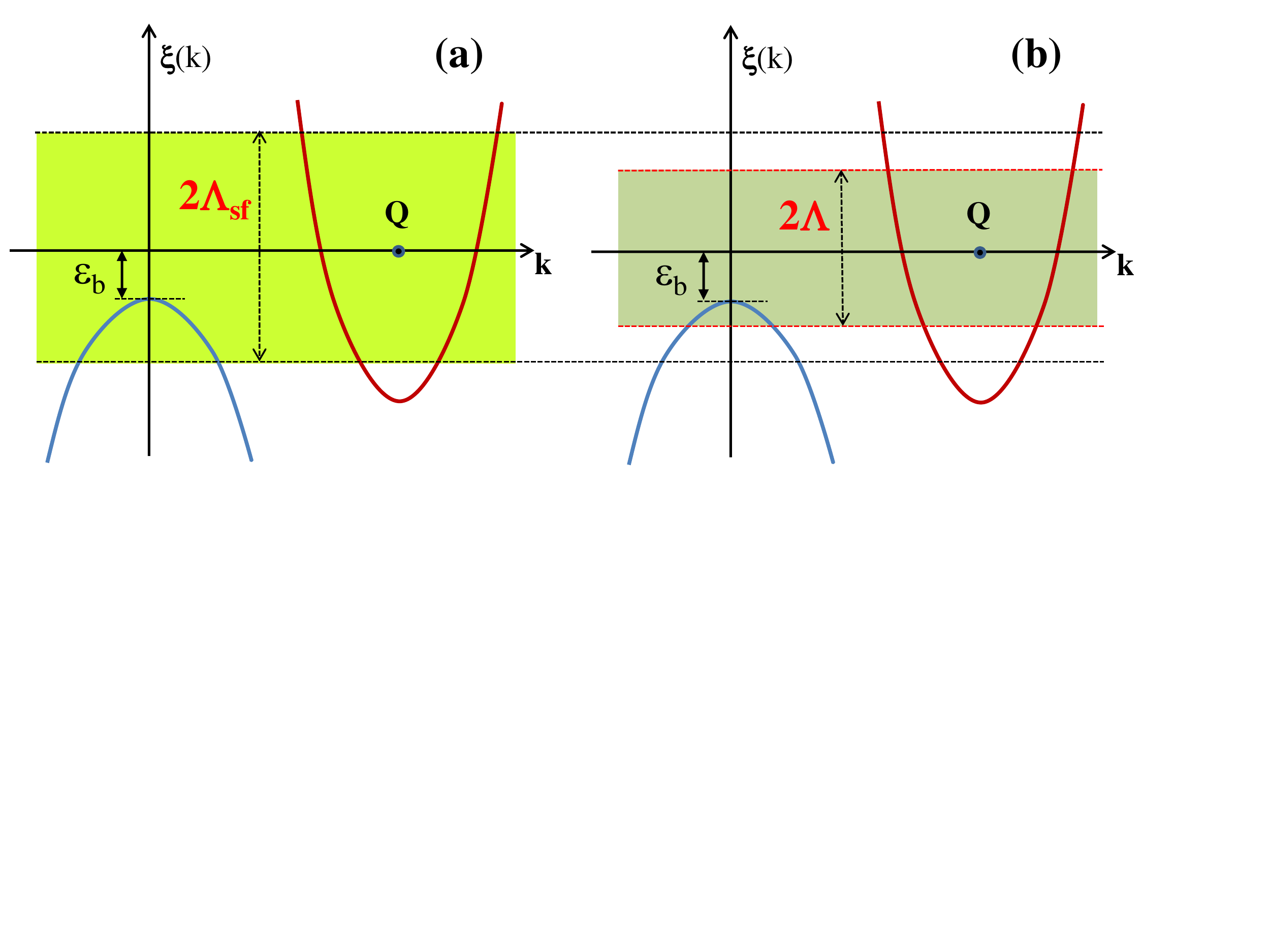}
\vspace{-4cm}
\caption{(Color online) (A) A typical incipient band model with $\Lambda_{sf} > \epsilon_b$.
(B) The same model but with the pairing cutoff scaled down as $\Lambda < \Lambda_{sf}$. For each value of $\Lambda$, $H_{ren}(\Lambda)$ is defined and the best SC gap solutions are calculated as: incipient $s_{he}^{-+}$-gap for $\epsilon_b < \Lambda < \Lambda_{sf}$, and $s_{ee}^{++}$-gap for $\Lambda <\epsilon_b$, with the same $T_c$ for all $\Lambda$.
\label{fig3}}
\end{figure}

In this paper, we assume the incipient energy $\epsilon_b$ to be smaller than the spin-fluctuation pairing interaction cutoff $\Lambda_{sf}$ as illustrated in Fig.3. This incipient band model\cite{Bang_shadow} and its extended models\cite{Hirschfeld_incipient} were recently studied as a possible candidate model for the HEDIS SC systems but only the incipient $s_{he}^{-+}$-type solution (Fig.2(a)) was investigated. Here we will consider the incipient $s_{he}^{-+}$-gap solution and $s^{++}_{ee}$-gap solution (Fig.2(b)) on an equal footing in the two band model. The $s^{+-}_{e1e2}$-gap solution (Fig.2(c)) will be considered with three band model later.

The coupled gap equations are the same as the ordinary two band SC model as follows.
\begin{eqnarray}
\Delta_h  &=&     \bigl[ V_{hh} \chi_h  \bigr] \Delta_h +
    \bigl[ V_{he}  \chi_e  \bigr] \Delta_e , \\ \nonumber
\Delta_e  &=&     \bigl[ V_{ee} \chi_e   \bigr] \Delta_e +
\bigl[ V_{eh} \chi_h  \bigr] \Delta_h
\end{eqnarray}
where the pair susceptibilities are defined as
\begin{eqnarray}
\chi_{h}(T) &=&  - \frac{N_h}{2} \int _{-\Lambda_{sf}} ^{-\epsilon_{b}} \frac{d\xi}{\xi}
\tanh (\frac{\xi}{2 T}) \approx - \frac{N_h}{2} \ln \Big[\frac{1.14 \Lambda_{sf}}{\epsilon_b}\Big],
\\ \nonumber
\chi_{e}(T) &=&  - N_e \int _{-\Lambda_{sf}} ^{\Lambda_{sf}} \frac{d\xi}{\xi}
\tanh (\frac{\xi}{2 T}) \approx - N_e \ln \Big[\frac{1.14 \Lambda_{sf}}{T}\Big]
\end{eqnarray}
where $N_{h,e}$ are the density of states (DOS) of the hole band and
electron band, respectively.
Assuming all repulsive pair potentials $V_{ab}$, but with $V_{inter-band} (=V_{he}, V_{eh}) > V_{intra-band} (=V_{hh}, V_{ee})$, the coupled gap equations Eq.(5) with the susceptibilities Eq.(6) produce the incipient $s^{+-}_{he}$-gap solution with the OPs $\Delta_h$ and $\Delta_e$ with opposite signs\cite{Bang_shadow,Hirschfeld_incipient}. $T_c$ decreases with increasing $\epsilon_b$; however the relative size of $|\Delta_h|/|\Delta_e|$ is insensitive to the value of $\epsilon_b$ and the gap size of the incipient hole band $|\Delta_h|$ is comparable to the size of $|\Delta_e|$.

\begin{figure}
\noindent
\includegraphics[width=110mm]{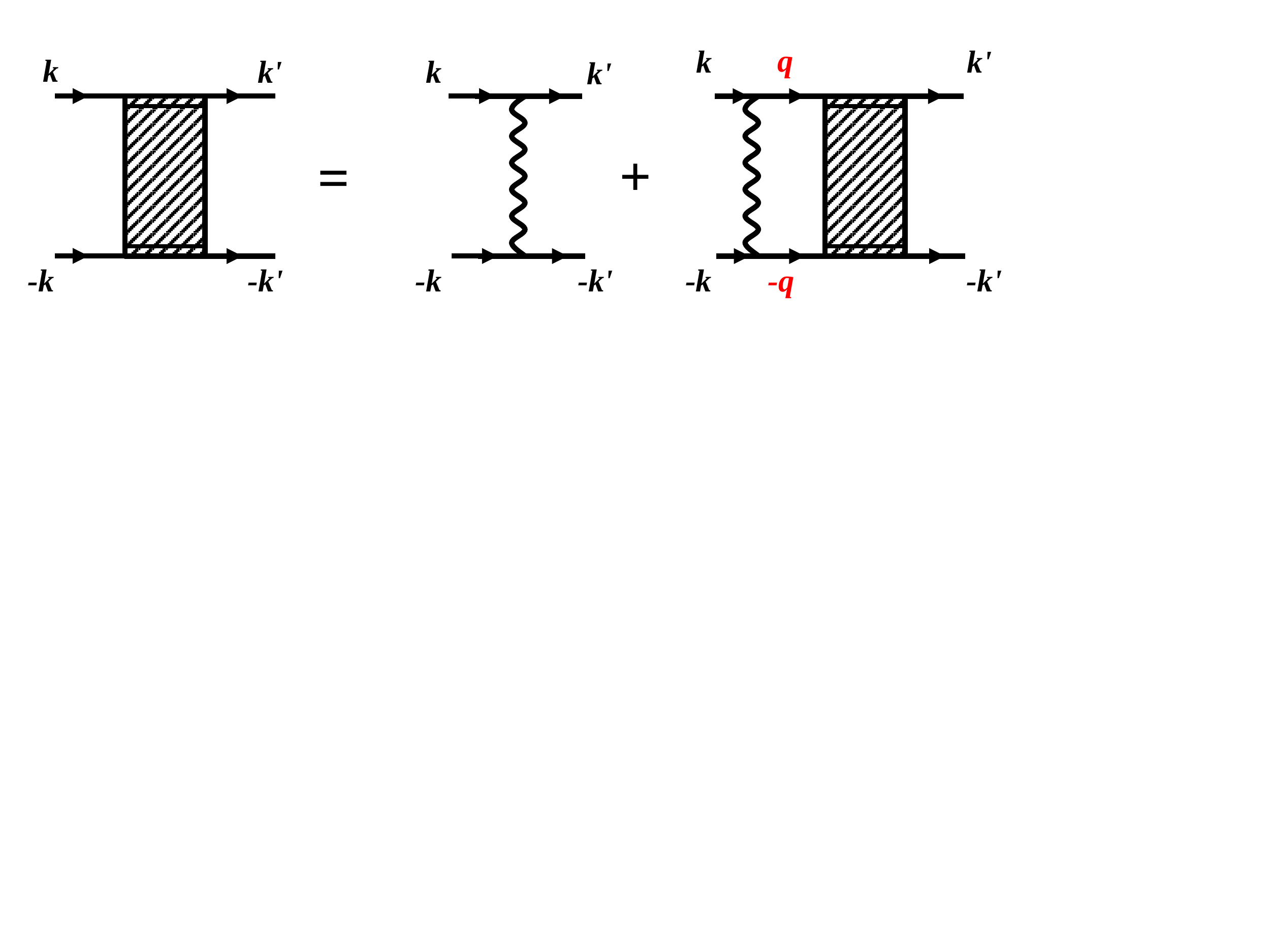}
\vspace{-6cm}
\caption{(Color online) Schematic diagram of the renormalization process. The lefthand side (hatched box vertex) is the renormalized pairing potential $\hat{V}_{ab}(\Lambda)$ defined at low energies ($ < \Lambda$) and the first term of the righthand side (the wiggly line vertex) is the bare pairing potential $\hat{V}_{ab}^{0}$. The momenta $\pm k, \pm k^{'}$ belong to the low energy region $|\xi(\pm k)|, |\xi(\pm k^{'})| \in [0:\Lambda]$.  The momenta $\pm q$ belong to the high energy region $|\xi(\pm q)|| \in [\Lambda:\Lambda_{sf}]$.
\label{fig4}}
\end{figure}

\begin{figure}
\noindent
\includegraphics[width=95mm]{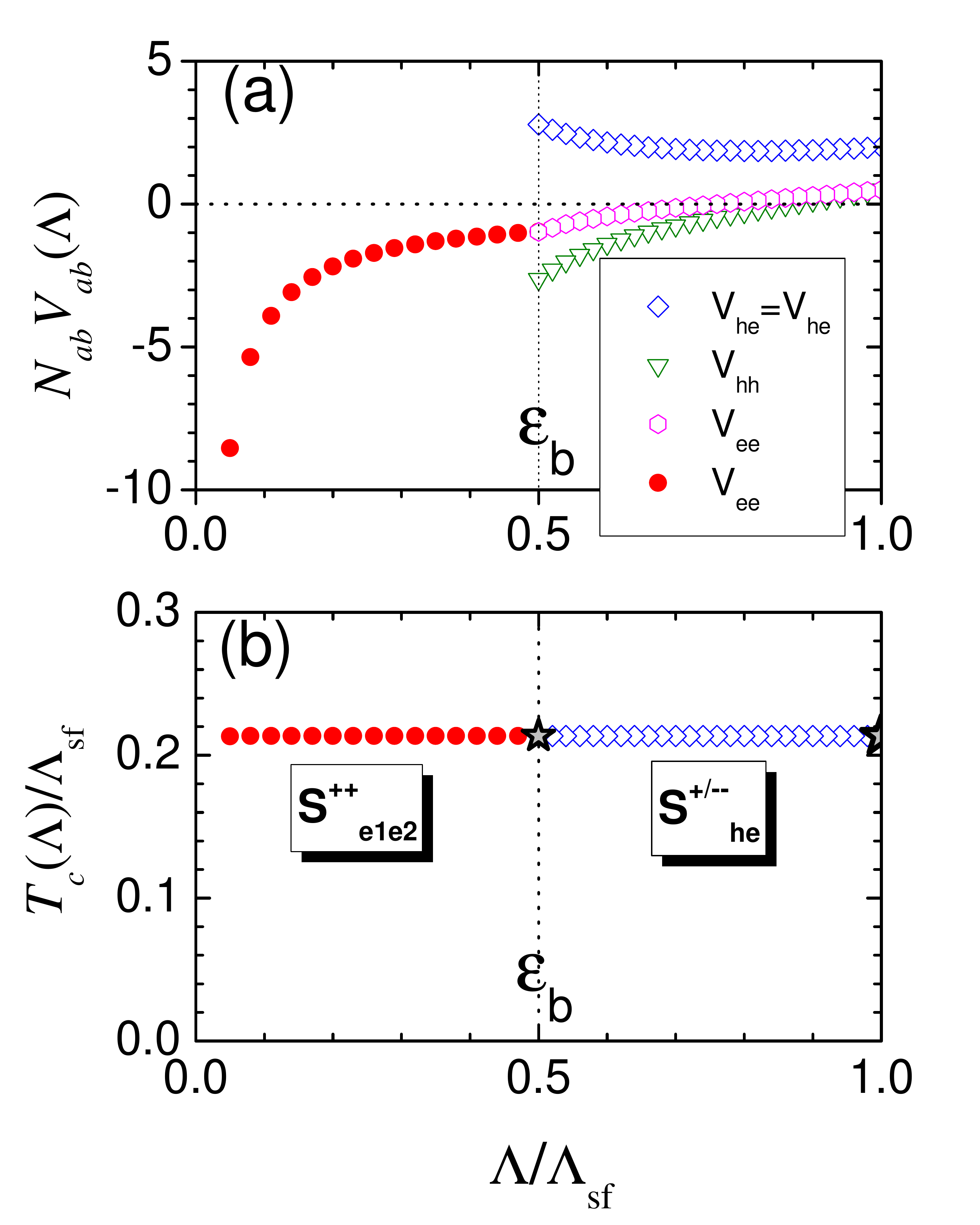}
\vspace{-0.5cm}
\caption{(Color online) Numerical results of the RG scaling of a typical incipient two band model of Fig.(3) with $\epsilon_b = 0.5 \Lambda_{sf}$. (A) Renormalized pair potentials $N_{ab}\hat{V}_{ab}(\Lambda)$ vs $\Lambda$ (with $N_{ab}=\sqrt{N_a N_b}$) with bare potentials: $N_hV^0_{hh}=N_eV^0_{ee}=0.5$ and $\sqrt{N_h N_e}V^0_{he}=\sqrt{N_h N_e}V^0_{eh}=2.0$ (B) Calculated $T_c$ with $\hat{V}_{ab}(\Lambda)$ and the corresponding pairing solutions. The pairing gap solution changes when $\Lambda$ crosses $\epsilon_b$, but $T_c$ remains the same all the time with scaling.
\label{fig5}}
\end{figure}

This model is already a low energy effective model in which all the high energy interaction processes originating from $U$ (Hubbard on-site interaction), $J$ (Hund coupling), etc are renormalized down to $\Lambda_{sf}$ to produce the effective pair potentials $V_{ab}^{0}(\Lambda_{sf})$ with cutoff energy $\Lambda_{sf}$. In particular, the characteristic AFM spin fluctuation energy scale $\Lambda_{sf}$ is an experimentally measurable -- by neutron experiments -- physical excitations just like a phonon energy $\omega_D$ in the BCS theory.
In a standard theory of superconductivity, the renormalization stops here and remained is to solve the gap equation(s) (e.g. Eq.(5)) by a mean field method (BCS theory) or by its dynamical extension (Eliashberg theory).

In this paper, we continue the RG scaling down to arbitrary low energy scale $\Lambda < \Lambda_{sf}$ and define the renormalized model $H_{ren}(\Lambda)$, as depicted in Fig.3(b).  The renormalized pair potential $V_{ab}(\Lambda)$ is defined by a standard RG process as follows (see Fig.4),
\begin{equation}
\hat{V}(\Lambda) =  \hat{V}^{0}+\hat{V}^{0}\cdot \hat{\chi}(\Lambda_{sf};\Lambda) \cdot \hat{V}(\Lambda)
\end{equation}
\noindent
and the formal solution is defined by
\begin{equation}
\hat{V}(\Lambda) =  [1 - \hat{V}^{0}\cdot \hat{\chi}(\Lambda_{sf};\Lambda)]^{-1} \cdot \hat{V}^{0},
\end{equation}
where $\hat{V}(\Lambda)$ is the renormalized 2x2 matrix pair potential with the new cutoff energy $\Lambda$, and  $\hat{V}^{0}$ is the bare pair potential with the cutoff energy $\Lambda_{sf}$ defined as
\begin{equation}
   \hat{V}^{0}=
  \left( \begin{array}{cc}
   V_{hh}^{0} & V_{he}^{0} \\      V_{eh}^{0} & V_{ee}^{0}  \end{array}  \right).
\end{equation}
Accordingly, the Cooper susceptibility $\hat{\chi}$ is also 2x2 matrix defined as
\begin{equation}
   \hat{\chi}(\Lambda_{sf};\Lambda)=
  \left( \begin{array}{cc}
   \chi_{h} & 0 \\      0 & \chi_{e}  \end{array}  \right)
\end{equation}
\noindent with $\chi_{e}(\Lambda_{sf};\Lambda)= -T_c\sum_n 2 N_{e}\int^{\Lambda_{sf}}_{\Lambda} d\xi \frac{1}{\omega_n^2 + \xi^2(k)}$ and $\chi_{h}(\Lambda_{sf};\Lambda)= - T_c\sum_n N_{h}\int_{-\Lambda_{sf}}^{-\Lambda} d\xi \frac{1}{\omega_n^2 + \xi^2(k)}$. Notice that the factor $2$ is missing in $\chi_{h}$ since the hole band exists only below the Fermi level. Also $\chi_{h}(\Lambda_{sf};\Lambda)$ is defined only up to $\Lambda \rightarrow \epsilon_b$, meaning that when the scaling cutoff $\Lambda$ runs below $\epsilon_b$ as $\Lambda < \epsilon_b$,  only the electron band will contribute to the RG flow.

In Fig.5(a), we show the results of the renormalized $\hat{V}_{ab}(\Lambda)$ for the whole range of $\Lambda <\Lambda_{sf}$. In this representative case, we chose $\epsilon_b=0.5 \Lambda_{sf}$ and the bare potentials: $N_hV^0_{hh}=N_eV^0_{ee}=0.5$ and $\sqrt{N_h N_e}V^0_{he}=\sqrt{N_h N_e}V^0_{eh}=2.0$. For $\epsilon_b < \Lambda <\Lambda_{sf}$, all four components of $\hat{V}_{ab}(\Lambda)$ get renormalized. The key features are: (1) the repulsive interband pair potentials $V_{he,eh}(\Lambda)$ slightly decrease at the beginning but eventually become more repulsive; this is very different behavior compared to the standard single band repulsive potential under RG such as the Coulomb potential. (2) the weaker repulsive intraband pair potentials $V_{hh,ee}(\Lambda)$ turn quickly into attractive ones; the different flow tracks of $V_{hh}(\Lambda)$ and $V_{ee}(\Lambda)$, despite the same starting bare potential $N_hV^0_{hh}=N_eV^0_{ee}=0.5$, is due to the difference of $\chi_{h}$ and $\chi_{e}$. For $\Lambda < \epsilon_b$, only $V_{ee}(\Lambda)$ continues to scale and other pair potentials stop scaling because $\chi_{h}$ stops scaling at $\Lambda = \epsilon_b$.

In Fig.5(b), we showed the calculated $T_c$ of the renormalized model $H_{ren}(\Lambda)$ for all values of $\Lambda < \Lambda_{sf}$ depicted in Fig.3(b). Namely, we solved the gap equations Eq.(5) in the limit of $\Delta_{h,e} \rightarrow 0$ with the renormalized pair potentials $V_{ab}(\Lambda)$ and the reduced cutoff $\Lambda$. For $\epsilon_b < \Lambda <\Lambda_{sf}$, the pairing gap solution is the incipient $s^{-+}_{he}$-solution, forming SC OPs both on the hole- ($\Delta^{-}$) and electron band ($\Delta^{+}$) (see Fig.2(a)). When $\Lambda < \epsilon_b$, the hole band is outside the pairing cutoff $\Lambda$, hence cannot participate forming Cooper pairs. Therefore the gap solution consists of the electron band only with attractive $V_{ee}(\Lambda)$, i.e. $s^{++}_{ee}$-solution (see Fig.2(b)).

Figure 5(b) shows that the $T_c (\Lambda)$ remain the same all the time with the scaling for $\Lambda <\Lambda_{sf}$. This is a consistent result with the invariance principle of RG\cite{note1} but still surprising to be confirmed with the multiband SC model with a complicated pair potential $\hat{V}_{ab}$, in particular, despite the change of the pairing solutions from $s^{-+}_{he}$ to $s^{++}_{ee}$\cite{note2}.
We would like to emphasize that, in order to achieve this $T_c$ invariance with RG transformation, it is crucially important to calculate the $T_c$-equation Eq.(5) together with the renormalized pair potentials $\hat{V}_{ab}(\Lambda;T=T_c)$ Eq.(8) at same temperature, which needs self-consistent iterative calculations of two equations, because the RG flow itself depends on temperature.

In fact, this electron band only pairing state, $s^{++}_{ee}$, has been suggested by previous works\cite{Wang2013, DHLee2011} of FRG studies, applied to the tight binding model with local interactions $U$, $U^{'}$, and $J_H$, designed for the FeSe systems. The FRG technique traces the RG flow of the effective pairing channels $V_{\alpha}(\Lambda) \phi_{\alpha}^{*}(k) \phi_{\alpha}(k')$ and identifies the most diverging channel "$\alpha$" as the winning instability of the system as the RG scale $\Lambda$ runs to arbitrarily low energy. However, because the $s^{++}_{ee}$-channel and the $s^{-+}_{he}$-channel are symmetry-wise identical and have the same form factor $\phi_{\alpha}(k) \sim (\cos{k_x}+\cos{k_y})$ (in two Fe/cell BZ), these two channels are running on the same track of the FRG flow. Therefore, identifying the most diverging form factor
$\phi_{\alpha}(k)$ itself does not distinguish which of these two pairing states is a true ground state.
What distinguishes these two pairing states is neither "channel" nor "symmetry" (they are always the same), but the physical pairing cutoff $\Lambda_{phys}$, however the FRG scheme itself does not determine the physical pairing cutoff $\Lambda_{phys}$.

This is exactly the point we are addressing in this paper: how to determine the physical pairing cutoff $\Lambda_{phys}$.
A common sense would be to identify as $\Lambda_{phys}=\Lambda_{sf}$ (the spin-fluctuation energy scale).
And when $\Lambda_{sf} > \epsilon_b$ (Fig.3(a)), the pairing solution should be the "incipient" $s^{-+}_{he}$-state forming gaps on the sunken hole band as well as on the electron bands\cite{Bang_shadow, Hirschfeld_incipient}.
And only if  $\Lambda_{sf} < \epsilon_b$ and $V^0_{ee}(\Lambda_{sf}) <0$, the pairing solution can be the $s^{++}_{ee}$-state, forming gaps only on the electron bands.
However, we have shown in Fig.5(b) that even when $\Lambda_{sf} > \epsilon_b$, there is no reason to fix the physical pairing cutoff as $\Lambda_{phys}=\Lambda_{sf}$ because the whole range of $\Lambda < \Lambda_{sf}$ produces the pairing states with the same $T_c$ and same symmetry but with different pairing cutoffs.

In the previous section, we have argued that among the degenerate pairing solutions with different pairing cutoffs $\Lambda$, the physical ground state should be chosen as the one with the largest $\Lambda$ because the condensation energy (CE) gain is larger with a larger cutoff energy $\Lambda$ when considering the higher order corrections $\sim O(T_c/\Lambda, \Delta_{h,e}/\Lambda)$ to the CE.
Therefore, among the continuously degenerate solutions of the $s^{\pm}_{he}(\Lambda)$-state for $\epsilon_b < \Lambda <\Lambda_{sf}$, the system should choose the incipient $s^{\pm}_{he}(\Lambda_{sf})$ state with cutoff energy, $\Lambda_{phys} =\Lambda_{sf}$, as a physical ground state. By the same reasoning, among the degenerate solutions of the $s^{++}_{ee}(\Lambda)$-state for $\Lambda <\epsilon_b$, the system will choose the $s^{++}_{ee}$ state with $\Lambda_{phys} = \epsilon_b$ as a physical ground state. These two physical ground states are marked as black star symbols in Fig.5(b) and specifically illustrated in Fig.6. Therefore, our work provides the rationale for justifying the electron band only pairing state, $s^{++}_{ee}$, with the physical cutoff $\Lambda_{phys}=\epsilon_b$ even when the original pairing cutoff is $\Lambda_{sf}$ that is larger than $\epsilon_b$.

\begin{figure}
\noindent
\includegraphics[width=95mm]{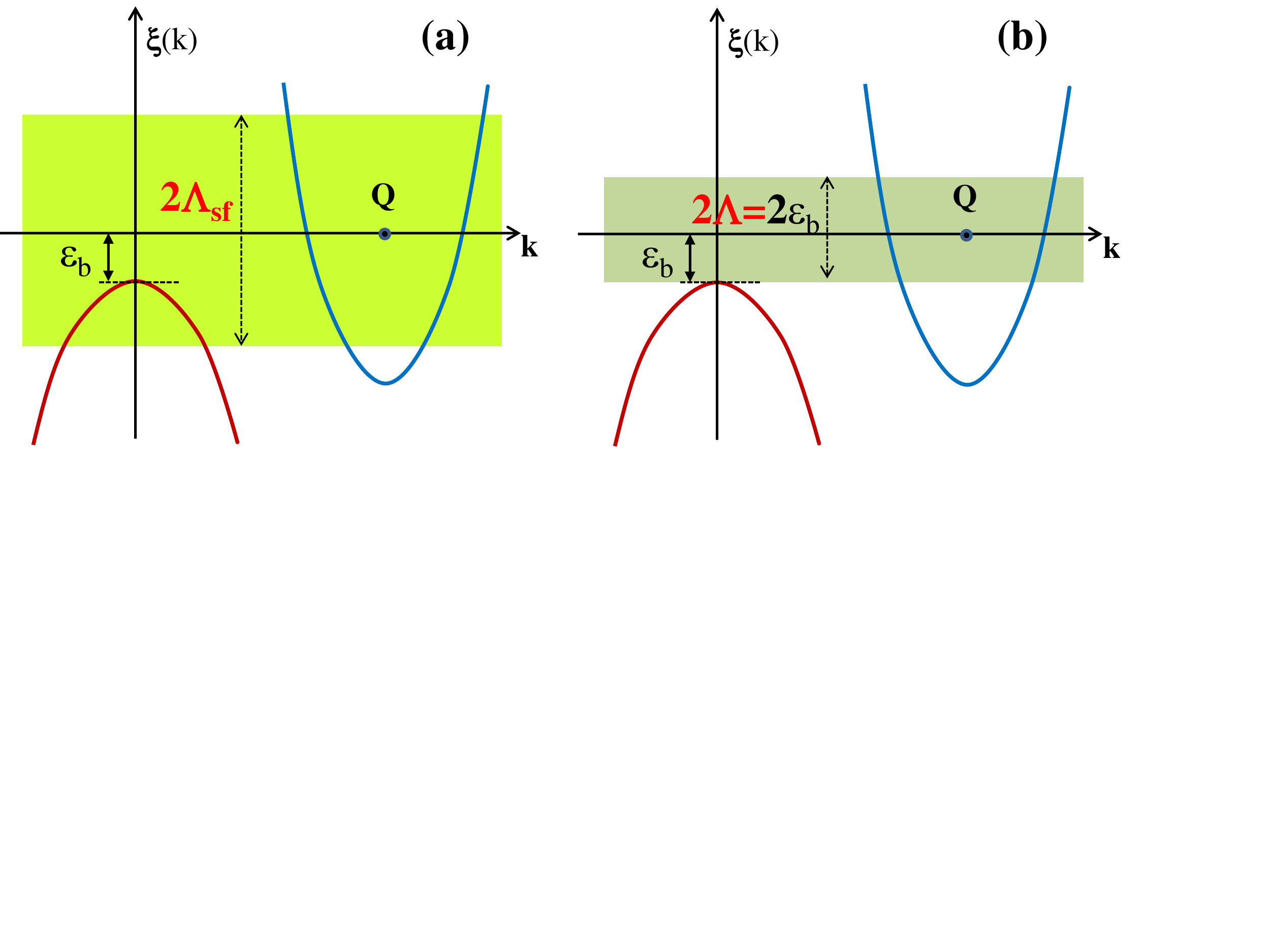}
\vspace{-4cm}
\caption{(Color online) (A) A typical incipient two band model with $\hat{V}^0_{ab}(\Lambda_{sf})$ and the pairing cutoff $\Lambda_{phys}=\Lambda_{sf}$; the pairing solution is the incipient $s_{he}^{\pm}$.
(B) The same model with the renormalized pairing potentials $\hat{V}_{ab}(\epsilon_b)$ and the reduced cutoff $\Lambda_{phys}=\epsilon_b$; the pairing solution is $s_{ee}^{++}$. The $T_c$ of both states are the same.
\label{fig6}}

\includegraphics[width=90mm]{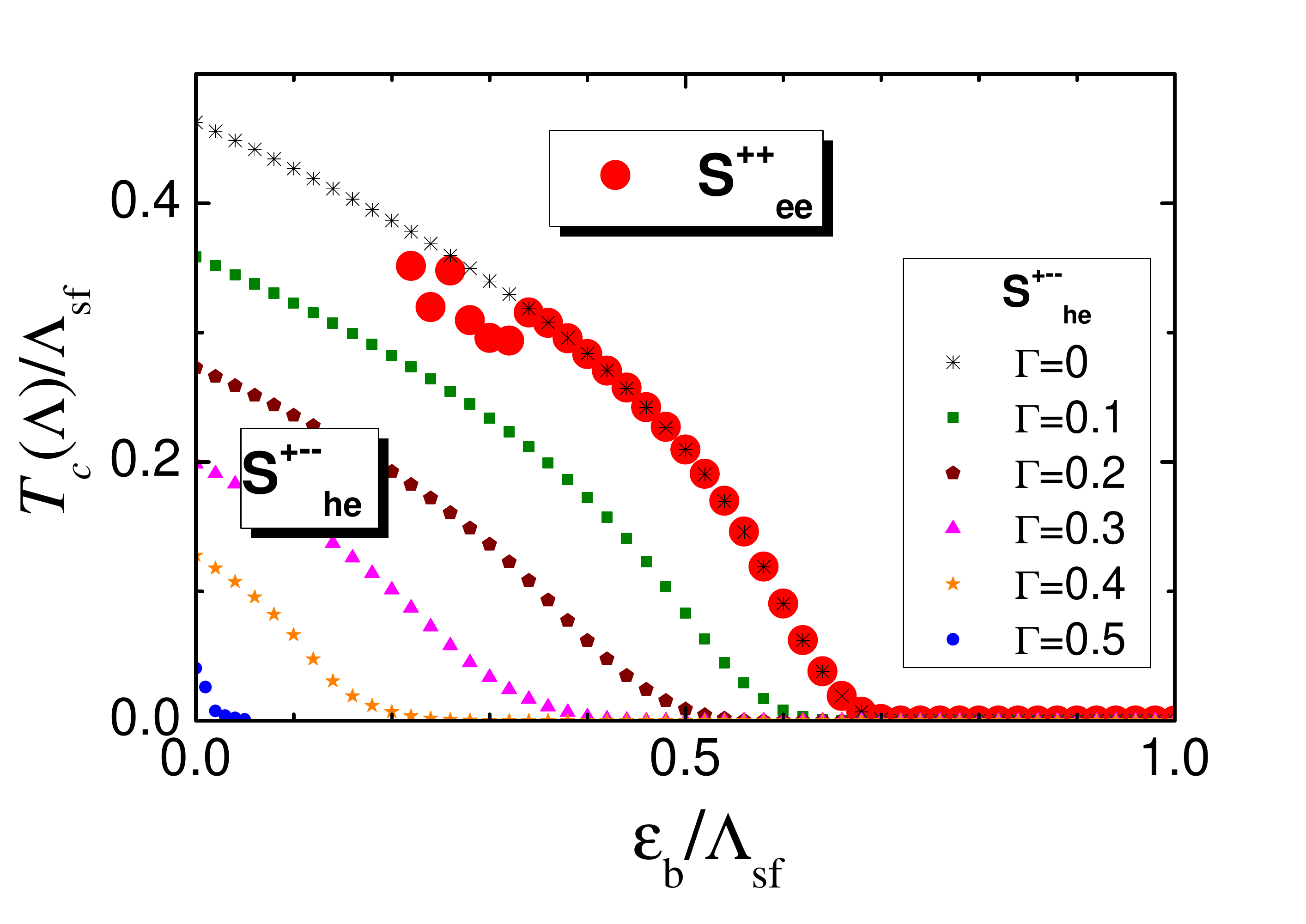}
\vspace{-0.5cm}
\caption{(Color online) Calculated $T_c$s as a function of $\epsilon_b$ for the two pairing states in Fig.6: the $s_{ee}^{++}$-state and the incipient $s_{he}^{\pm}$-state with impurity pair-breaking rate $\Gamma/\Lambda_{sf}=0, 0.1, 0.2, 0.3, 0.4$ and 0.5, respectively. The $T_c$s of $s_{he}^{\pm}$-state are systematically suppressed with increasing $\Gamma$, but $T_c$s of $s_{ee}^{++}$-state (red solid circles) do not change with impurities. Pairing potentials are the same as in Fig.5: $N_hV^0_{hh}=N_eV^0_{ee}=0.5$ and $\sqrt{N_h N_e}V^0_{he}=\sqrt{N_h N_e}V^0_{eh}=2.0$
\label{fig7}}
\end{figure}

\subsection{Incipient $s^{\pm}_{he}$-state, $s^{++}_{ee}$-states, and impurity scattering}

Now we compare two physical ground state solutions of a given incipient two band model with $\Lambda_{sf} > \epsilon_b$: the incipient $s^{\pm}_{he}$-state with $\Lambda_{phys} =\Lambda_{sf}$ and the $s^{++}_{ee}$-state with $\Lambda_{phys} = \epsilon_b$ (see Fig.6). We numerically calculate the $T_c$ of these two gap solutions with varying $\epsilon_b$ for $0 < \epsilon_b < \Lambda_{sf}$.
For all these calculations, it is important to calculate the Cooper susceptibilities without approximation because the cutoff energy $\Lambda_{phys}=\epsilon_b$ of the $s^{++}_{ee}$-state can be very low to violate the BCS limit ($\frac{T_c}{\Lambda_{phys}} \ll 1$). Therefore, we numerically calculate $\chi_{e}(\Lambda;0)= -2T\sum_n N_{e}\int^{\Lambda}_{0} d\xi \frac{1}{\omega_n^2 + \xi^2(k)}$, which can be very different from the BCS approximation $\chi_{e}(\Lambda;0) \sim -N_e \ln{\frac{1.14\Lambda}{T}}$ when $\Lambda < T$.
We chose the same interaction parameters, $N_hV^0_{hh}=N_eV^0_{ee}=0.5$ and $\sqrt{N_h N_e}V^0_{he}=\sqrt{N_h N_e}V^0_{eh}=2.0$ as in the case of Fig.5.

Figure.7 shows that the $T_c$s of the incipient $s^{\pm}_{he}$-state with $\Lambda_{phys}=\Lambda_{sf}$ (black stars) and the $T_c$s of the $s^{++}_{ee}$-state with $\Lambda_{phys}=\epsilon_b$ (red solid circles) are exactly overlayed on top of each other, in clean limit ($\Gamma=0$), as explained before. $T_c$ decreases with increasing $\epsilon_b$ as the hole band sinks deeper below Fermi level. One important new feature is that while the $T_c$ of the incipient $s^{\pm}_{he}$-state continuously increases as $\epsilon_b \rightarrow 0$, the $T_c$ of the $s^{++}_{ee}$-state becomes ill-defined when $\epsilon_b  \rightarrow 0$. This is because the Cooper susceptibility $\chi_{e}(\Lambda=\epsilon_b;T)$ gets saturated and loses its logarithmic divergence as $\epsilon_b \sim O(T_c)$; in fact $\chi_{e}(\Lambda=\epsilon_b;T)$ decreases, instead of increasing, with lowering temperature when $\epsilon_b \sim O(T_c)$ so that the $T_c$-equation fails to find a solution. This behavior provides an important clue why there exists a minimum incipient distance energy $\epsilon_b (=\Lambda_{mim})$ below which the $s^{++}_{ee}$-pairing state is not stabilized regardless of the strength of the pair potentials $\hat{V}_{ab}(\Lambda)$. In clean limit, this minimum cutoff energy scale is $\Lambda_{mim} \sim T_c$, but it will increase with additional relaxation processes, thermal or dynamical origins, such as $\Lambda_{mim} \sim (\pi T_c + \Gamma_{imp} +\Gamma_{inela})$, where $\Gamma_{imp}$ is static impurity scattering rate, and $\Gamma_{inelas}$ is inelastic scattering rate that can be provided by spin fluctuations as $\Gamma_{inelas}=Im \Sigma_{sf}(T)$.

We now consider the impurity pair-breaking effect on $T_c$ of both pairing states and in this paper we considered non-magnetic impurities only. First, we have to investigate the impurity effect on the RG scaling itself. The answer is that the non-magnetic impurities has no effect on the RG scaling because the Cooperon propagators (pair susceptibilities) $\chi_{h,e}(\Lambda_{sf};\Lambda)$ defined in Eq.(10) are invariant with the non-magnetic impurity scattering. The fundamental reason for this invariance has the same origin as the Anderson's theorem\cite{Anderson_imp}: the $s$-wave pairing is not affected by the non-magnetic impurity scattering. This is easy to understand by noting that the Cooperon propagators $\chi_{h,e}(\Lambda_{sf};\Lambda)$ entering the RG equation are nothing but the $s$-wave pair susceptibility. Diagrammatic derivation of this proof following the formalism of Abrikosov and Gor'kov\cite{AG} is given in Appendix A. Therefore, the renormalized pair potentials $\hat{V}(\Lambda)$ are the same with and without the non-magnetic impurities.

Then given the same pairing potentials $\hat{V}(\Lambda)$, the non-magnetic impurity scattering would not change $T_c$ of the s-wave pairing according to Anderson's theorem\cite{Anderson_imp}, so that the $T_c$ of the $s^{++}_{ee}$-state in Fig.7 will not be affected with non-magnetic impurities. On the other hand, it is well known that non-magnetic impurities will suppress the $T_c$ of the "standard" $s^{\pm}_{he}$-state almost as strong as in the d-wave\cite{Bang_imp}. We might expect that the $T_c$ suppression with non-magnetic impurities on the "incipient" $s^{\pm}_{he}$-state will be weakened compared to the case of the standard $s^{\pm}_{he}$-state because the incipient band, being sunken below Fermi level, should be less effective for any scattering process.
However, we have found that this weakening effect due to the incipiency is only marginal and the non-magnetic impurity $T_c$-suppression rate of the "incipient" $s^{\pm}_{he}$-state is almost as strong as the case of the "standard" $s^{\pm}_{he}$-state. The physical reason is because the relative size of OPs, $\Delta_h$ and $\Delta_e$ are not affected by the incipient distance energy $\epsilon_b$ as far as the pairing interactions is dominated by the interband potentials as $V_{he,eh} > V_{hh,ee}$\cite{Bang_shadow, Hirschfeld_incipient}. The detailed formalism of the impurity effect on the incipient $s^{\pm}_{he}$-state is described in Appendix B.

In Fig.7, we plotted the results of $T_c$ suppression of the incipient $s^{\pm}_{he}$-state with the different impurity pair-breaking rate $\Gamma/\Lambda_{sf}=0.1, 0.2, 0.3, 0.4$, and $0.5$, respectively. In this paper, we used the equal strength unitary scatterers ($c=0$) both for inter- and intra-band impurity scatterings as $V_{he,eh}^{imp}=V_{hh,ee}^{imp}$ (see Appendix A).
As expected, the $T_c$ suppression rate is strong and comparable to the case of a standard $s^{\pm}_{he}$-state.
The key messages of Fig.7 is:\\
(1) The $T_c$s of two degenerate pairing solutions, the incipient $s^{\pm}_{he}$-state and the $s^{++}_{ee}$-state, of the incipient two band model tract each other in clean limit. \\
(2) When non-magnetic impurities exist, this degeneracy breaks down and the incipient $s^{\pm}_{he}$-pairing state quickly becomes suppressed with impurities, while the $s^{++}_{ee}$-state is robust against the non-magnetic impurity pair-breaking.\\
(3) Most interestingly, however, the $s^{++}_{ee}$-state cannot be stabilized when the incipient hole band approaches too close to Fermi level such as $\epsilon_b  < (\pi T_c + \Gamma_{imp} +\Gamma_{inela})$. This implies that there exists an optimal incipient energy distance $\epsilon_b^{optimal}$ for the $s^{++}_{ee}$-pairing state.

\vspace{0cm}
\begin{figure}
\noindent
\includegraphics[width=90mm]{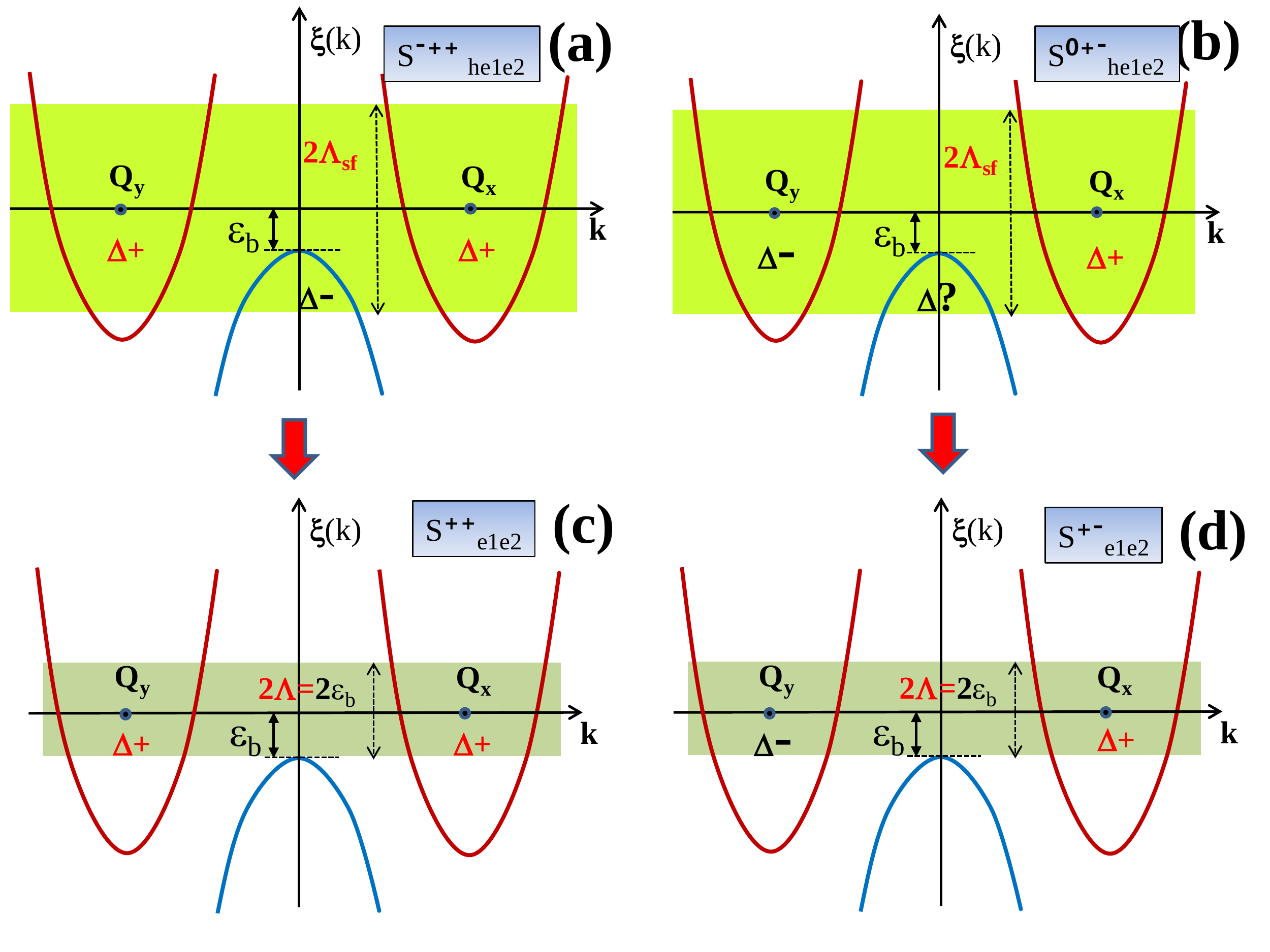}
\caption{(Color online) Schematic pictures of possible pairing solutions of the three band model.
(a), (b) Solutions without RG scaling but with original pairing cutoff $\Lambda_{sf}$. (c), (d) Solutions with the renormalized pairing cutoff $\Lambda_{phys}=\epsilon_b$. Through the RG flow, (a) flows to (c) with the same $T_c^{(1)}$, and (b) flows to (d) with another same $T_c^{(2)}$, respectively, but crossings are not possible. \label{fig8}}
\end{figure}

\subsection{Incipient three band model}
\begin{figure}
\noindent
\includegraphics[width=90mm]{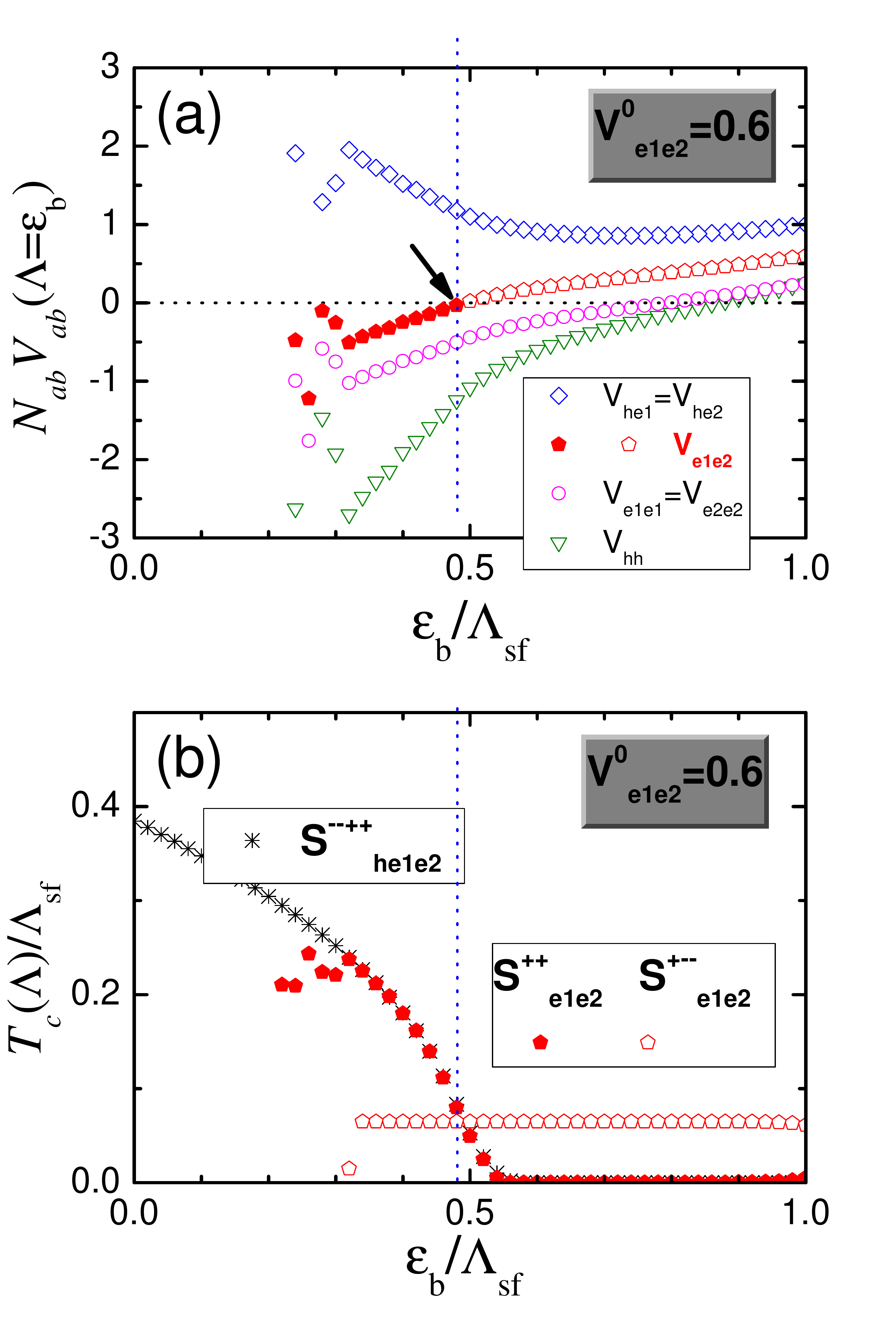}
\vspace{-0.5cm}
\caption{(Color online) Numerical results of the RG scaling of a typical incipient three band model of Fig.(8). (a) Renormalized pair potentials $N_{ab}\hat{V}_{ab}(\Lambda)$ vs $\Lambda=\epsilon_b$ ($N_{ab}=\sqrt{N_a N_b}$) of the three band model with bare pairing potentials: $\sqrt{N_h N_e}V^0_{he}=1.0$, $\sqrt{N_{e1} N_{e2}}V^0_{e1e2}=0.6$, $N_h V^0_{hh}=0.25$ and $N_e V^0_{ee}=0.25$. When the renormalized  $N_{e1e2}V_{e1e2}(\Lambda)$ (red pentagons) becomes attractive ($\epsilon_b/\Lambda_{sf} < 0.48$; denoted by black arrow), the maximum $T_c$ solution changes from $s_{e1e2}^{+-}$ (open pentagons) to $s_{e1e2}^{++}$ (solid pentagons).  (b) Calculated $T_c$s vs $\epsilon_b$ of the pairing solutions illustrated in Fig.8.
\label{fig9}}
\end{figure}
Now we consider more realistic three band model: one incipient hole band ($h$), plus two electron bands ($e1, e2$) (see Fig.8). In particular, we can study the effect of the inter-electron band interaction $V_{e1,e2}$ which might have a non-negligible strength when the magnetic correlation has a deviation from the standard $C$-type (${\bf Q}=(\pi,0), (0,\pi)$) toward $G$-type (${\bf Q}=(\pi,\pi)$) \cite{G-type,G-type2,G-type3}. If the $G$-type magnetic correlation is dominant, i.e., when $V_{e1,e2} > V_{he1}, V_{he2}$, the leading pairing solution should always be the one in which $\Delta_{e1}$ and $\Delta_{e2}$ have the opposite signs but the same sizes (Fig.2(c), Fig.8(b), Fig.8(d)) \cite{DHLee2011,d-wave1,d-wave2,d-wave3}. This gap solution is also called as "nodeless $d$-wave" or "$d_{x^2-y^2}$", etc. in the literature, but in this paper we denote it as "$s^{+-}_{e1e2}$" to be contrasted to "$s^{++}_{e1e2}$-state".
In real FeSe systems, it is most likely that the C-type correlation is dominant, but mixed with some fraction of the G-type correlation\cite{C-type, G-type,G-type2,G-type3}. Therefore, in the three band model studied in this paper, we assumed $V_{e1e2} < V_{he} (=V_{he1}=V_{he2})$ along with the assumption of all repulsive ($V_{ab} >0$) inter- and intra-band pair potentials. In this case, we found that there is an interesting transition from the $s^{+-}_{e1e2}$-state to the $s^{++}_{e1e2}$-state as the incipient energy $\epsilon_b$ varies.

In Fig.8, we have sketched the typical band structure of the three band model with its possible pairing solutions. As in two band model, we assumed $\Lambda_{sf} > \epsilon_b$. In Fig.8(a) and (b), without RG scaling, two possible pairing solutions, $s^{-++}_{he1e2}$ and $s^{0+-}_{he1e2}$, are illustrated. The $s^{-++}_{he1e2}$-state is the same state as the incipient $s^{-+}_{he}$-state of the two band model in the previous section. The $s^{0+-}_{he1e2}$-state in Fig.8(b) is subtle. First, we found that this pairing state with OPs $\Delta_{e1}^{+}$ and $\Delta_{e2}^{-}$, with opposite signs each other on the electron bands $e1$ and $e2$, is possible even with $V_{e1,e2} < V_{he1}, V_{he2}$, as will be shown with numerical calculations. Second, having the OPs $\Delta_{e1}^{+}$ and $\Delta_{e2}^{-}$ with opposite signs, the OP on incipient hole band $\Delta_{h}$ can be anything: positive, negative, or zero, without altering the pairing energy with the pair potentials $V_{he1}$ and $V_{he2}$. But when considering $V_{hh} >0$, $\Delta_{h}=0$ is the best solution. However, this state is still not exactly the same state as the $s^{+-}_{e1e2}$-state in Fig.8(d) because they have different physical pairing cutoffs, $\Lambda_{phys}=\Lambda_{sf}$ and $\Lambda_{phys}= \epsilon_b$, respectively.
In Fig.8(c) and (d), two obviously possible solutions, $s^{++}_{e1e2}$ and $s^{+-}_{e1e2}$, with electron bands only are illustrated. As indicated by the vertical red arrows, the RG flows are: (a) to (c), and (b) to (d), but crossing flow between them are not possible. Therefore the $T_c$s of (a) and (c) are equal and the $T_c$s of (b) and (d) are equal, respectively.

Figure9 shows the numerical results of a representative case of the three band model with the dimensionless bare pairing potentials $N_{ab}V^0_{ab}=\sqrt{N_a N_b}V^0_{ab}$: $\sqrt{N_h N_e}V^0_{he}=1.0$, $\sqrt{N_{e1} N_{e2}}V^0_{e1e2}=0.6$, $N_h V^0_{hh}=0.25$ and $N_e V^0_{ee}=0.25$, where we assumed $N_{e1}=N_{e2}$, $V^0_{ee}=V^0_{e1e1}=V^0_{e2e2}$, and $V^0_{he}=V^0_{he1}=V^0_{he2}$. In Fig.9(a), the renormalized potentials $N_{ab}V_{ab}(\Lambda=\epsilon_b)$ are plotted as functions of $\epsilon_b$. These potentials are used to calculate $T_c$s of the $s^{++}_{e1e2}$ and $s^{+-}_{e1e2}$ states illustrated in Fig.8(c) and (d) with the physical cutoff $\Lambda_{phys}=\epsilon_b$. The results of $T_c$ versus $\epsilon_b$ are plotted in Fig.9(b): $s^{++}_{e1e2}$ (solid red pentagons) and $s^{+-}_{e1e2}$ (open red pentagons), respectively. when the incipient hole band is deep (for large $\epsilon_b$), $s^{+-}_{e1e2}$-state (nodeless $d$-wave) has the highest $T_c$, and as the incipient hole band becomes shallow (for smaller $\epsilon_b$), $s^{++}_{e1e2}$-state becomes winning. This transition happens when the renormalized inter-electron band potential $V_{e1e2}(\epsilon_b)$ turns into negative as indicated by black arrow in Fig.9(a).

Interestingly, Fig.9(b) shows that the $T_c$ of the $s^{+-}_{e1e2}$-state (nodeless $d$-wave; open red pentagons) doesn't change as $\epsilon_b$ varies. This behavior can be understood, however, if we remember that the $T_c$ of Fig.8(b) and the $T_c$ of Fig.8(d) should be the same due to the RG invariance. In the pairing state of Fig.8(b), i.e. $s^{0+-}_{he1e2}$-state, we argued $\Delta_h =0$. Then it is easy to note that with the given bare values of $N_{ab}V^0_{ab}$ and $\Lambda_{sf}$, varying $\epsilon_b$ has no effect on the gap equation and $T_c$. On the other hand, in the cases of $s^{-++}_{he1e2}$ (Fig.8(a)) and $s^{++}_{e1e2}$ (Fig.8(c)) -- they have the same $T_c$ in clean limit -- varying $\epsilon_b$ should strongly affect $T_c$ because $\Delta_h \neq 0$ in the $s^{-++}_{he1e2}$-state. Fig.9(b) shows this behavior of $T_c$ versus $\epsilon_b$ for the $s^{-++}_{he1e2}$ (black stars) and $s^{++}_{e1e2}$ (solid red pentagons) states. As in the case of two band model, the $T_c$s of the incipient $s^{-++}_{he1e2}$-state with the fixed pairing cutoff energy $\Lambda_{phys}=\Lambda_{sf}$ continuously increases as $\epsilon_b \rightarrow 0$. However, the $T_c$s of the $s^{++}_{e1e2}$-state with the pairing cutoff energy $\Lambda_{phys}=\epsilon_b$ becomes ill-defined when $\epsilon_b < T_c$.

\subsection{Impurity effect on three band model}
Now we would like to consider the non-magnetic impurity effect on the $T_c$ of $s^{-++}_{he1e2}$-, $s^{++}_{e1e2}$-, and $s^{+-}_{e1e2}$-states. As we have argued in the two band model, the $T_c$ of the $s^{++}_{e1e2}$-state is immune to the non-magnetic impurity scattering. The impurity effect on the  $s^{+-}_{e1e2}$-state (nodeless $d$-wave) is mathematically equivalent to the case of the $d$-wave state, hence we expect the strongest $T_c$ suppression. Finally, the impurity effect on the incipient $s^{-++}_{he1e2}$-state is the same to the case of the incipient $s^{-+}_{he}$-state in two band model. The impurity theory for the two band model can be straight forwardly used for the three band model with a replacement of $N_e = N_{e1}+N_{e2}$ (see Appendix A).

\begin{figure}
\noindent
\includegraphics[width=90mm]{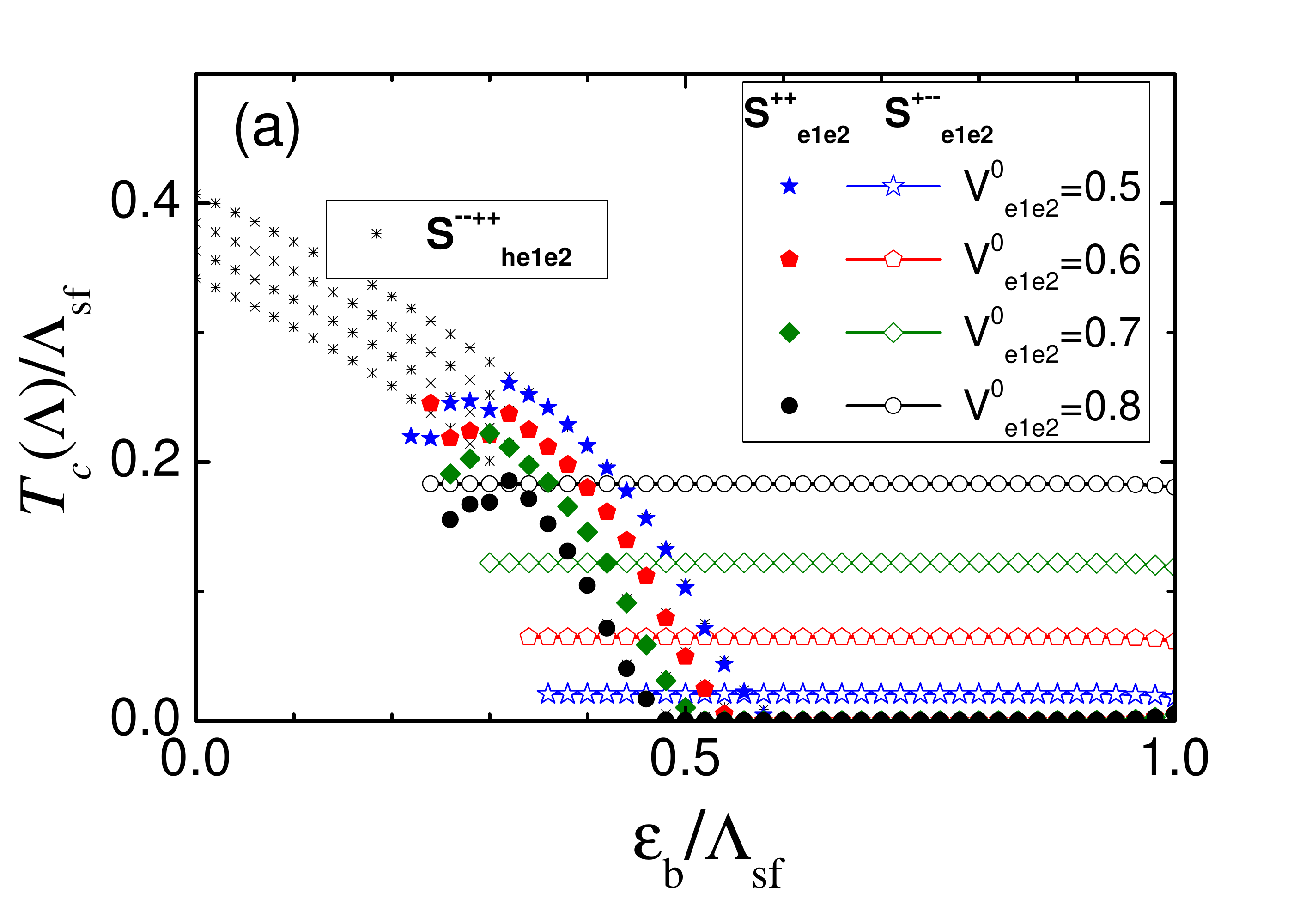}
\includegraphics[width=90mm]{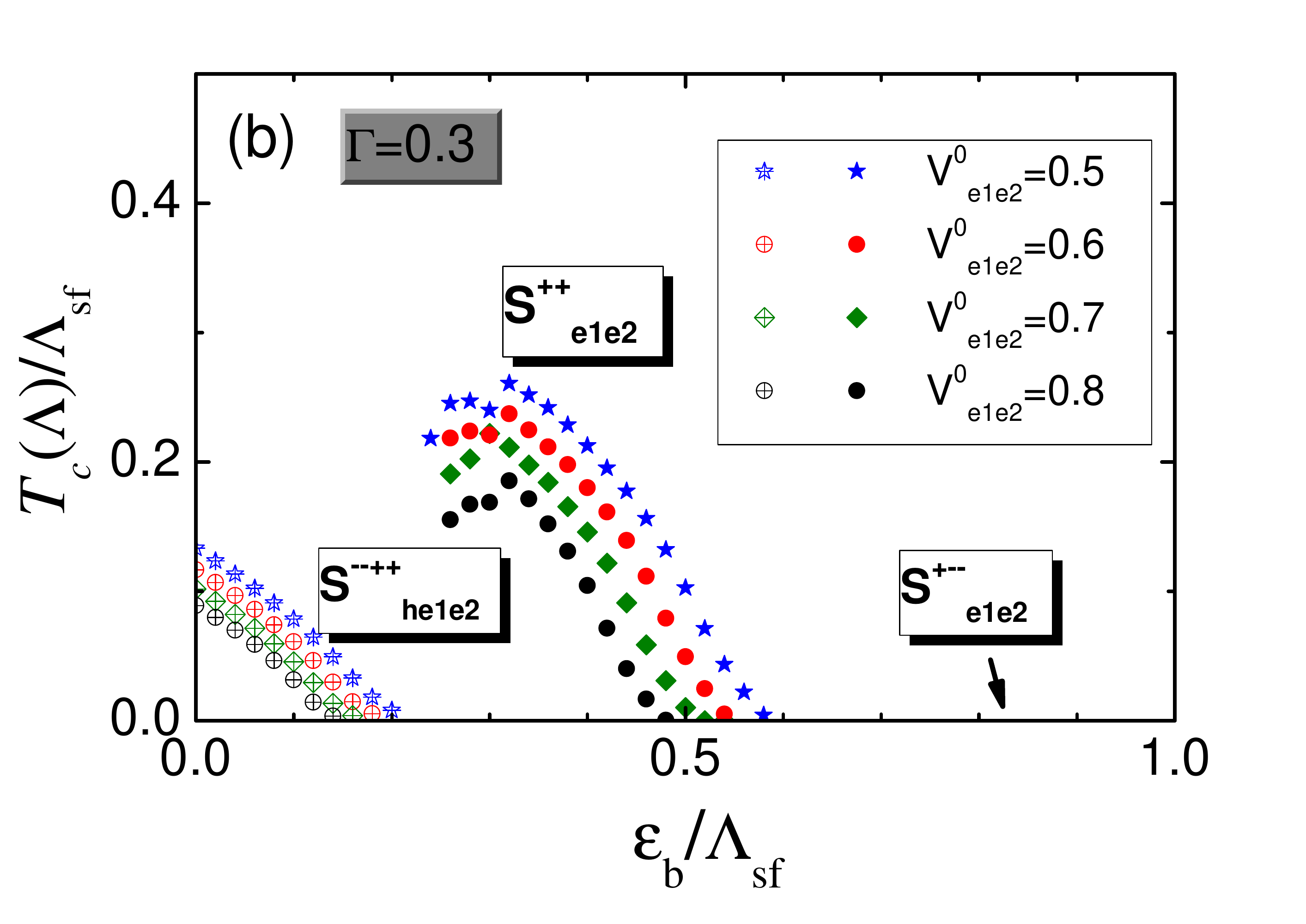}
\vspace{0cm}
\caption{(Color online) (a) Calculated $T_c$s vs $\epsilon_b$ of the pairing solutions illustrated in Fig.8. for different bare inter-electronband potential $\sqrt{N_{e1} N_{e2}}V^0_{e1e2}=0.5, 0.6, 0.7$, and 0.8, respectively. Other bare pairing potentials are $\sqrt{N_h N_e}V^0_{he}=1.0$, $N_h V^0_{hh}=0.25$ and $N_e V^0_{ee}=0.25$ (b) The same calculations as (a) but with non-magnetic impurity scattering rate $\Gamma/\Lambda_{sf}=0.3$.
\label{fig10}}
\end{figure}

Figure10(a) is the same plot as Fig.9(b) -- $T_c$ versus $\epsilon_b$ for three pairing states: $s^{-++}_{he1e2}$, $s^{++}_{e1e2}$, and $s^{+-}_{e1e2}$ -- with the same parameters as in Fig.9 except for varying values of $N_{e1e2}V^0_{e1e2}=0.5, 0.6, 0.7,$ and 0.8, respectively.
It is busy plot but easy to understand the general trend. First, the $T_c$ of the $s^{+-}_{e1e2}$-state (nodeless $d$-wave; open symbols) increases with increasing $N_{e1e2}V^0_{e1e2}$. On the contrary, the $T_c$s of the $s^{-++}_{he1e2}$ (black stars) and $s^{++}_{e1e2}$-state (solid symbols) decreases with increasing $N_{e1e2}V^0_{e1e2}$, as expected. Secondly, as already shown in Fig.9(b), there are crossovers of the higher $T_c$ pairing state as $\epsilon_b$ decreases from the $s^{+-}_{e1e2}$-state (nodeless $d$-wave; open symbols) to the $s^{++}_{e1e2}$-state (solid symbols) for each value of $V^0_{e1e2}$. Finally, the $T_c$s of the $s^{++}_{e1e2}$-state is ill-defined when $\epsilon_b < T_c$.

Figure10(b) shows the same calculations of Figure10(a) but including non-magnetic impurity scattering.
We assumed unitary limit scatterers ($c$=0; see Appendix A) for all inter- and intra-band scatterings and the impurity scattering rate $\Gamma=0.3\Lambda_{sf}$. This is quite a large scattering rate and this value was chosen to kill all $T_c$ of the $s^{+-}_{e1e2}$-state (nodeless $d$-wave) just for illustration. While the $T_c$s of the $s^{++}_{e1e2}$-state remain the same as in Fig.10(a), the $T_c$s of the incipient $s^{-++}_{he1e2}$-state are strongly suppressed. As a result, the incipient $s^{-++}_{he1e2}$-state (open symbols) survives only in the region of the left corner of small values of $\epsilon_b$ in Fig.10(b).
The results of Fig.10(b) imply the following simple picture. If we can change the depth of the incipient hole band, $\epsilon_b$, by electron doping, pressure, or dosing \cite{double_dome1,double_dome2,double_dome3,FeSe_dosing,FeSe_dosing2} in a typical FeSe system, the shallow incipient hole band system (small $\epsilon_b$) can support the incipient $s^{-++}_{he1e2}$-state with low $T_c$, while the $s^{++}_{e1e2}$-state can appear with much higher $T_c$ with increasing $\epsilon_b$ to the optimal value $\epsilon_b^{optimal}$. Further increasing impurity scattering rate, all incipient $s^{-++}_{he1e2}$-state disappears and only the $s^{++}_{e1e2}$-state will survive with high $T_c$ in the region of $\epsilon_b^{optimal}$.
Incidently, these results of Fig.10(b) looks very similar to the recent experimental observations of the phase diagram of electron doped FeSe systems, which shows the curious double dome and single dome structure of the $T_c$ versus electron doping phase diagram\cite{double_dome1,double_dome2,double_dome3}.

\section{Summary and Conclusions}
We believe that the FeSe/STO monolayer system is an exception among all HEDIS system, in that it has an extra phonon boost effect to achieve the exceptionally high $T_c$ up to $\sim$100K\cite{FeSe3}. In this paper, we focused on the common electronic pairing mechanism with the electron pockets only, the characteristics shared by all HEDIS systems including the FeSe/STO monolayer system.
We studied the pairing mechanism of the phenomenological incipient band models: one incipient hole band ($h$) plus one electron band ($e$) or two electron bands ($e1,e2$) with the pairing interactions $V_{ab}^0 >0$ -- possibly provided by the AFM spin-fluctuations -- with the original pairing cutoff energy $\Lambda_{sf} (> \epsilon_b)$.

We introduced the concept of {\it dynamical tuning of physical cutoff by RG}. Using this concept and direct numerical calculations, we found that the incipient band model allows two degenerate SC solutions with the exactly same $T_c$ in clean limit: the $s^{\pm}_{he}$-gap ($\Delta_h^{-} \neq 0$, $\Delta_e^{+} \neq 0$) and $s_{ee}^{++}$-gap ($\Delta_h =0$, $\Delta_e^{+} \neq 0$) solutions with different pairing cutoffs, $\Lambda_{phys}=\Lambda_{sf}$ and $\Lambda_{phys}=\epsilon_b$, respectively.
The $s_{ee}^{++}$-gap solution, with $\Lambda_{phys}=\epsilon_b$, actively eliminates the incipient hole band from forming Cooper pairs, and becomes immune to the impurity pair-breaking.
As a result, the HEDIS systems, by dynamically tuning the pairing cutoff and by selecting the $s_{ee}^{++}$-pairing state, can always achieve the maximum $T_c$ -- the $T_c$ of the degenerate $s^{\pm}_{he}$ solution in the ideal clean limit -- latent in the original pairing interactions, even in dirty limit.
We also found that there exist an optimal incipient energy $\epsilon_b^{optimal}$ of the hole band, below this value the $s_{ee}^{++}$-pairing state cannot be stabilized.  We estimated $\epsilon_b^{optimal}  \approx (\pi T_c + \Gamma_{imp} + \Gamma_{inelas})$.

With more realistic three band model with one incipient hole band ($h$) and two electron bands ($e1, e2$), we  also considered additional pairing state: $s^{+-}_{e1e2}$, also called as "nodeless $d$-wave" state.
We showed in general that the $s^{+-}_{e1e2}$-state is favored when the incipient hole band is deep (large $\epsilon_b$) but the $s^{++}_{e1e2}$-state becomes favored when the incipient hole band becomes intermediate to shallow depth ($\epsilon_b \sim \epsilon_b^{optimal}$). Including non-magnetic impurity scattering, the $s^{+-}_{e1e2}$-state (nodeless $d$-wave) becomes most rapidly destroyed and the "incipient" $s^{-++}_{he1e2}$-state might barely survive with very low $T_c$ in the region of small values of $\epsilon_b$. However, the $s^{++}_{e1e2}$-state, being immune to the non-magnetic impurity pair-breaking, can exist with much higher $T_c$ in the region of optimal values of $\epsilon_b^{optimal}$. This double-dome structure of the phase diagram shown in Fig.10(b) and the general trend of the transition of the pairing states: "incipient" $s^{-++}_{he1e2}$-state $\rightarrow$  $s^{++}_{e1e2}$-state, with electron doping, accompanied with increasing $T_c$, looks very much similar to the recent experiments of electron doped FeSe systems\cite{double_dome1,double_dome2,double_dome3,FeSe_dosing, FeSe_dosing2}.

In conclusion, we showed: (1) The standard paradigm of the IBS superconductivity\cite{Hirschfeld,Hirschfeld2} -- the $s^{\pm}_{he}$-pairing mediated by a dominant interband repulsion between the hole band(s) around $\Gamma$ and the electron band(s) around $M$ -- continues to operate in the HEDIS systems.  (2) The new ingredient of the pairing mechanism in the HEDIS systems is the {\it dynamical tuning of the pairing cutoff $\Lambda_{phys}$ from $\Lambda_{sf}$ to $\epsilon_b$ by RG scaling}, which allows the $s^{++}_{e1e2}$-pairing state as the ground state of the HEDIS systems. By this, the HEDIS systems can avoid the impurity pair-breaking effect. But the drawback is that the hole band has to sink below Fermi level by $\epsilon_b$, which has to reduce $T_c$.
In view of this picture and relatively high $T_c \sim 30K-40$K of the HEDIS, all pnictide and chalcogenide IBS seem to suffer severe $T_c$-suppression due to intrinsic impurities, inevitably introduced with dopings; otherwise, the IBS systems in general could have had much higher $T_c$.
(3) The FeSe/STO monolayer system is special among other HEDIS systems, which has the additional {\it phonon boost effect}\cite{Bang_phonon1,Bang_phonon2,FeSe_phonon,DHLee_phonon,DHLee_phonon2,Johnston,Johnston2} on top of the above mentioned common pairing mechanism of the HEDIS systems.

{\it Acknowledgements -- } This work was supported by NRF Grant
2013-R1A1A2-057535 funded by the National Research Foundation of
Korea.

\appendix
\numberwithin{figure}{section}
\numberwithin{equation}{section}

\section{RG scaling with non-magnetic impurities}
Here we prove that the RG scaling in general is not affected by the non-magnetic impurity scattering. The effect of impurity scattering enters the Cooperon propagators at $T_c$ defined in Eq.(10) in the main text as follows.
\begin{eqnarray}
\tilde{\chi}_{e}(\Lambda_{sf};\Lambda) &=& -T_c\sum_n 2 N_{e}\int^{\Lambda_{sf}}_{\Lambda} d\xi \frac{\eta_v}{\tilde{\omega}_n^2 + \xi^2}, \\
\tilde{\chi}_{h}(\Lambda_{sf};\Lambda) &=& - T_c\sum_n N_{h}\int_{-\Lambda_{sf}}^{-\Lambda} d\xi \frac{\eta_v}{\tilde{\omega}_n^2 + \xi^2}.
\end{eqnarray}
\noindent
where $\tilde{\omega}_n=\omega_n+\Sigma_{imp}(\omega_n)$ is the renormalized Matsubara frequency by the non-magnetic impurity scattering, which can be explicitly calculated as $\tilde{\omega}_n=\omega_n \eta_1$ with $\eta_1=1+1/2\tau_1|\omega_n|$\cite{AG}. The process of $\Sigma_{imp}(\omega_n)$ is depicted in Fig.A.1(b). And $\eta_v$ is the corresponding vertex correction as depicted in Fig.A.1(a). Abrikosov and Gor'kov\cite{AG} has shown that $\eta_v = \eta_1$ exactly in the case of the non-magnetic impurity scattering and that the above formulas of the renormalized Cooperon propagators $\tilde{\chi}_{e,h}(T)$ become the same as the bare Cooperon propagators $\chi_{e,h}(T)$. Therefore the RG scaling of the renormalized potentials $\hat{V}(\Lambda)$ in Eq.(8) in the main text is not affected by the non-magnetic impurity scattering. Q.E.D.

\begin{figure}
\noindent
\includegraphics[width=100mm]{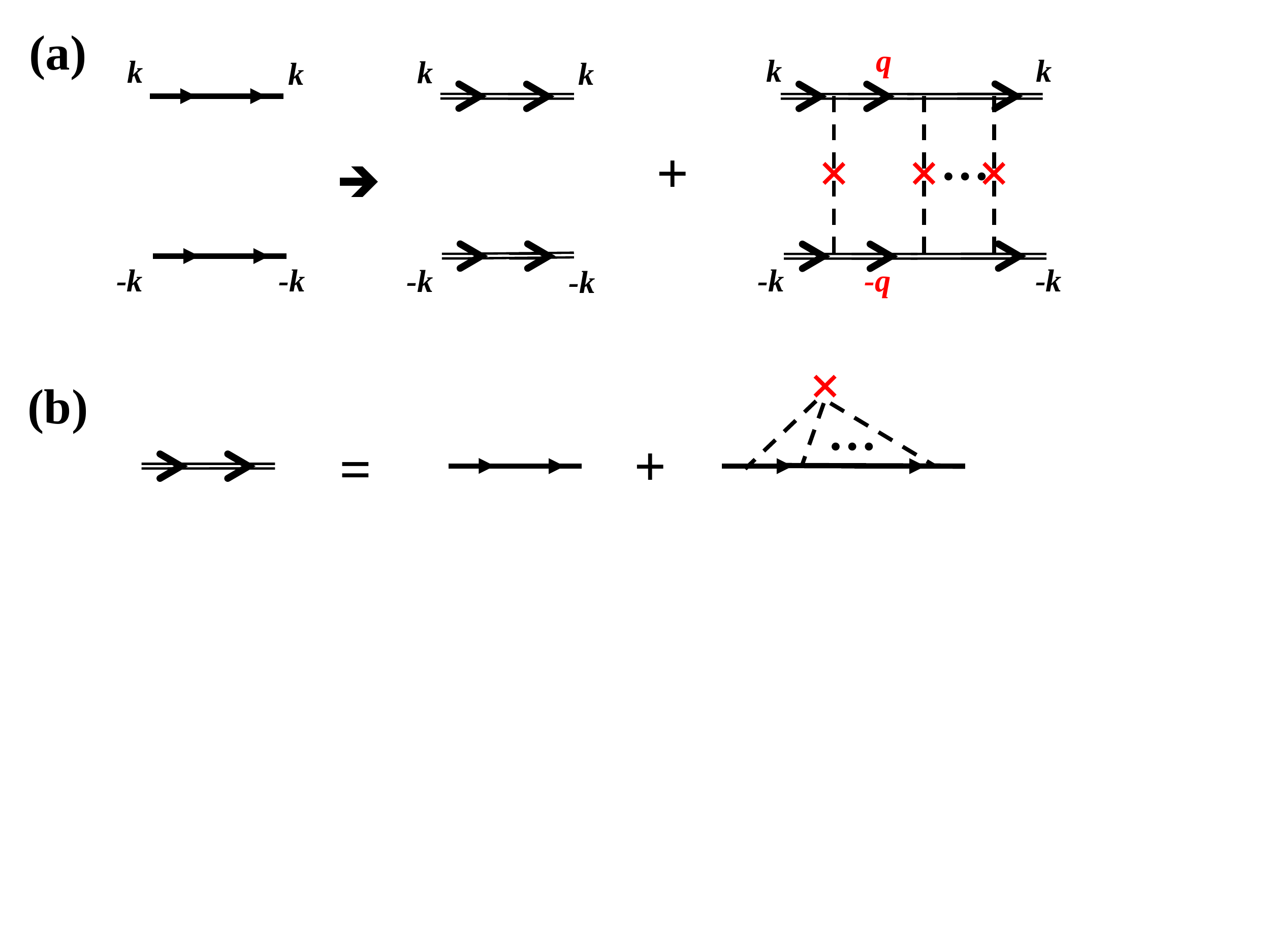}
\vspace{-4cm}
\caption{(Color online) (a) A bare Cooperon propagator $\chi(T)$ (lefthand side) is modified to be the renormalized one $\tilde{\chi}(T)$ (righthand side) by multiple impurity scatterings. (b) The renormalized fermion propagator $\tilde{G}$ (double line) by multiple impurity scatterings. All red crosses mean the non-magnetic impurity potential of the same single impurity site.
\label{figA1}}
\end{figure}
\section{Impurity Formalism of $T_c$-suppression}

\subsection{The two band incipient $s_{he}^{-+}$-state}
The impurity effect enters the pair susceptibilities $\tilde{\chi}(T)_{h,e}$ for $T < T_c$ in the gap equations Eq.(5) in the main text as follows,

\begin{eqnarray}
\tilde{\chi}_{h}(T) &=& -T \sum _n N_{h} \int _{-\Lambda_{sf}}
^{-\epsilon_b} d \xi \frac{ \tilde{\Delta}_{h}(k) } {
\tilde{\omega}_n^2 +\xi^2 + \tilde{\Delta}_{h} ^2 (k)},\\
\tilde{\chi}_{e}(T) &=& -T \sum _n N_{e} \int _{-\Lambda_{sf}}
^{\Lambda_{sf}} d \xi \frac{ \tilde{\Delta}_{e}(k) } {
\tilde{\omega}_n^2 +\xi^2 + \tilde{\Delta}_{e} ^2 (k)},
\end{eqnarray}
where $\tilde{\omega}_n$ and $\tilde{\Delta}_{h,e}$ contain all selfenergy corrections due to the impurity scattering. Notice the difference of the quasiparticle energy integration domains, $\int _{-\Lambda_{sf}}^{-\epsilon_b}$ and $\int _{-\Lambda_{sf}}^{\Lambda_{sf}}$, between the hole and electron band, respectively. For a standard $s_{he}^{-+}$-state with ordinary hole- and electron-band, where both bands has the integration domains of $\int _{-\Lambda_{sf}}^{\Lambda_{sf}}$, $\tilde{\omega}_n$ and $\tilde{\Delta}_{h,e}$ can be calculated using the $\mathcal{T}$-matrix method as\cite{Bang_imp}

\ba \tilde{\omega}_n =\omega_n + \Sigma^{0}
 _h(\omega_n) + \Sigma^{0} _e(\omega_n),  \\
\tilde{\Delta}_{h,e} = \Delta_{h,e} + \Sigma^1 _{h} (\omega_n) +
\Sigma^1 _{e} (\omega_n),  \\
\Sigma_{h,e} ^{0,1} (\omega_n)  = \Gamma \cdot \mathcal{T}^{0,1}
_{h,e} (\omega_n), ~~ \Gamma= \frac{n_{imp}}{\pi N_{tot}},
 \ea
where $\omega_n= T \pi (2n +1)$ is the Matsubara frequency,
$n_{imp}$ the impurity concentration, and $N_{tot}=N_h +N_e$
is the total DOS. The $\mathcal{T}$-matrices $\mathcal{T}^{0,1}$
are the Pauli matrices $\tau^{0,1}$ components in the Nambu space, and are written
for two band superconductivity as

\ba
\mathcal{T}^{i} _{a} (\omega_n) &=& \frac{G^{i} _{a} (\omega_n)}{D} ~~~~~(i=0,1; ~~a=h,e), \\
 D &=& c^2 +[G^0 _h + G^0 _e]^2 + [G^1 _h + G^1 _e]^2,\\
G^0 _a (\omega_n) &=& \frac{N_a}{N_{tot}}
\frac{\tilde{\omega}_n}
{\sqrt{\tilde{\omega}_n^2 + \tilde{\Delta}_{a} ^2 (k) }},\\
G^1 _a (\omega_n) &=& \frac{N_a}{N_{tot}} \frac{\tilde{\Delta}_{a}} {\sqrt{\tilde{\omega}_n^2 +
\tilde{\Delta}_{a} ^2 (k) }}, \ea
where $c=\cot\delta_0=1/[\pi N_{tot}V_{imp}]$ is a convenient measure of scattering strength and we assumed the equal strength for both inter- and intra-band impurity scatterings.
$c=0$ means the unitary limit ($V_{imp} \rightarrow \infty$) and $c > 1$ is the Born limit ($V_{imp} << 1$) scattering.

Now we need a modification of the above formulas for the incipient hole band model.
For $T_c$-equation ($\Delta_{h,e} \rightarrow 0$ limit), all differences come from the restricted integration of the incipient hole band below Fermi level as
\begin{equation}
N_{h} \int _{-\Lambda_{sf}}
^{-\epsilon_b} d \xi \frac{1} {\omega_n^2 +\xi^2} \approx N_h \frac{1}{|\omega_n|} \tan^{-1}{(\frac{\Lambda_{sf}}{\epsilon_b})},
\end{equation}

\noindent compared to the standard two band case with the ordinary hole band,
\begin{equation}
N_{h} \int _{-\Lambda_{sf}}
^{\Lambda_{sf}} d \xi \frac{1} {\omega_n^2 +\xi^2} \approx N_h \frac{\pi}{|\omega_n|}.
\end{equation}

\noindent Therefore, by replacing $N_h$ by $N_h^{eff}=N_h \tan^{-1}{(\frac{\Lambda_{sf}}{\epsilon_b})}/\pi$, all the above formulas for the standard two band model can be continuously used without further changes. We only need to redefine: $N_{tot}=(N_h^{eff}+N_e)$, and $\tilde{N}_h=N_h^{eff}/N_{tot}$, $\tilde{N}_e=N_e/N_{tot}$, and the strength of the impurity scattering rate $\Gamma= \frac{n_{imp}}{\pi N_{tot}}$ should be understood with newly defined $N_{tot}=(N_h^{eff}+N_e)$.
Being $N_h^{eff} < N_h$, it succinctly captures all the effects arising from the sunken hole band by $\epsilon_b$.

To determine $T_c$, we take $T \rightarrow T_c$ limit and
linearize the gap equations (5) in the main text with respect to the OPs
$\Delta_{h,e}$. First, the impurity renormalized Matsubara
frequency Eq.(B.3) and the OPs Eq.(B.4) are written as

\ba \tilde{\omega}_n
&=& \omega_n (1+ \eta_{\omega}), \\
\tilde{\Delta}_{h,e} &=& \Delta_{h,e} (1+\delta_{h,e}),
 \ea
with
 \ba
\eta_{\omega} &=& \frac{\Gamma}{1+c^2} \frac{1}{|\omega_n|},  \\
\delta_{h,e} &=& \frac{\Gamma}{1+c^2} \frac{1}{|\omega_n|}
\frac{[\tilde{N_h}\Delta_{h}+ \tilde{N_e}\Delta_{e}]}
{\Delta_{h,e}}.
 \ea
And the pair susceptibilities Eq.(B.1) and Eq.(B.2) are now simplified as
\ba \label{chi_he} \tilde{\chi}_{h,e}(T) = - \pi T N_{tot} \tilde{N}_{h,e}\sum _n \frac{
\Delta_{h,e}(k) (1+\delta_{h,e}) } {|\omega _n (1+
\eta_{\omega})|}.
\ea
\noindent
It is immediately clear that if $\eta_{\omega}=\delta_{a}$, as in a
single band $s$-wave state, there is no renormalization of
the pair susceptibility $\tilde{\chi}_{a}(k)$ with the impurity
scattering. This is just the Anderson theorem of $T_c$-invariance of the
$s$-wave SC. For a $d$-wave case, obviously $\delta_a=0$ and $\eta_{\omega} \neq 0$, hence results in the maximum $T_c$-suppression. In the case of a standard $s_{he}^{+-}$-wave, it was shown that $|\delta_a| \approx 0$ because of the inverse relation of $\frac{N_h}{N_e} \approx \frac{|\Delta_e|}{|\Delta_h|}$ and Eq.(B.15), in the limit of the dominant interband pairing ($V_{he,eh} >> V_{hh,ee}$)\cite{Bang_imp}.

In our incipient two band case, it is more complicated to draw
a simple conclusion. However, as we have shown above, after replacing $N_h \rightarrow N_h^{eff}$, all the formulas are the same as in the case of the standard $s_{he}^{+-}$-wave state. The remaining question is whether the inverse relation $\frac{N_h^{eff}}{N_e} \approx \frac{|\Delta_e|}{|\Delta_h|}$ is still hold or not, and we found that this relation is still hold in the incipient $s_{he}^{+-}$-state with numerical calculations\cite{Bang_shadow,Hirschfeld_incipient}. With the susceptibilities Eq.(B.16) together with the gap equations (5) in the main text in the limit $\Delta_a \rightarrow 0$, we have calculated $T_c$ in Fig.7 with non-magnetic impurity scattering. For convenience of parametrization, we used the impurity scattering rate parameter $\Gamma=n_{imp}/(\pi N_{tot})$ with $N_{tot}=(N_h+N_e)$ in Fig.7 instead of using the physically more relevant parameter $\Gamma=n_{imp}/(\pi N_{tot})$ with $N_{tot}=(N_h^{eff}+N_e)$, which is a complicate function of $\epsilon_b$.

\subsection{The three band incipient $s_{he1e2}^{-++}$-state}
Noticing that $N_{e1}=N_{e2}$ and $\Delta_{e1}=\Delta_{e2}$ (see Fig.2(a)), the above formulas for the two band incipient $s_{he}^{-+}$-case can be used only with the replacement of $N_e=2 N_{e1, e2}$. And the result of $T_c$-suppression is qualitatively the same as the incipient two band  $s_{he}^{+-}$-state.

\subsection{The three band $s_{e1e2}^{-+}$-state}
Noticing that this is a $d$-wave (nodeless) state (see Fig.2(c)), it is always $\delta_{a=e1,e2}=0$, hence results in the maximum $T_c$-suppression as in a standard $d$-wave case.

\subsection{The three band $s_{e1e2}^{++}$-state}
Finally, the most focused pairing state of this paper, the three band $s_{e1e2}^{++}$-state, having $\Delta_{e1}=\Delta_{e2}$ with the same sign and $\Delta_{h}=0$, this state is the same as the single band $s$-wave state, hence should satisfy the Anderson's theorem\cite{Anderson_imp} of the $T_c$-invariance with the time-reversal invariant disorders such as the non-magnetic impurities.


\begin{references}
\bibitem{note1}
The simple but important principle of RG transformation is that the physically measurable quantities should be the same before and after RG transformation. In this sense, we report a rare case in this paper in which this RG invariance is subtly modified, where although the pairing symmetry and $T_c$ remain invariant with RG transformation, the ground state can change with different values of physical pairing cutoff $\Lambda_{phys}$.
\bibitem{note2}
We emphasize that although the symmetry of the incipient  $s^{-+}_{he}$- and $s^{++}_{ee}$-state is the same, they are physically distinct states having different physical cutoff $\Lambda_{phys}$. For example, the hole band of the incipient  $s^{-+}_{he}$-state has the Bogoliubov quasiparticles and shadow gap\cite{Bang_shadow}, but the hole band of the $s^{++}_{ee}$-state remains normal state.
\end{references}
\end{document}